%% file: mlgameoflife.tex
\documentclass{article}

\usepackage[T1]{fontenc} 
\usepackage[latin1]{inputenc}
\usepackage{aeguill}
\usepackage{tikz}
\usepackage{amsfonts}
\usepackage{amsmath}
\usepackage{amssymb}
\usepackage{gnuplot-lua-tikz}
\usepackage{subfigure}
\usepackage[]{url}
\usepackage[ruled,linesnumbered]{algorithm2e}

\usetikzlibrary{arrows,automata,shapes}
\usepackage{tikz-uml}

\newenvironment{keywords}{
       \list{}{\advance\topsep by0.35cm\relax\small
       \leftmargin=1cm
       \labelwidth=0.35cm
       \listparindent=0.35cm
       \itemindent\listparindent
       \rightmargin\leftmargin}\item[\hskip\labelsep
                                     \bfseries Keywords:]}
     {\endlist}

\begin{document}

\title{Multi-level agent-based modeling\\with the Influence Reaction principle}

\author{Gildas Morvan, Daniel Jolly\\~\\
Univ. Lille Nord de France\\1bis rue Georges Lefèvre 59044 Lille cedex, France\\~\\LGI2A, U. Artois, Technoparc Futura 62400 Béthune, France.\\~\\ email: first name.surname@univ-artois.fr}

\date{}

\maketitle

\begin{abstract}
This paper deals with the specification and the implementation of multi-level agent-based models, using a formal model, IRM4MLS  (an Influence Reaction Model for Multi-Level Simulation), based on the Influence Reaction principle. Proposed examples illustrate forms of top-down control in (multi-level) multi-agent based-simulations.
\end{abstract}


\section{Introduction}

\setcounter{footnote}{0}

Four main \textit{theoretical} issues emerge in the literature on multi-level\footnote{The term multi-scale may also be found. Intuitively, a level and scale are similar concepts that both mean viewpoint. However, this notion should be clarified. In the following we assume that two agents are not at the same \textit{scale} iff they represent processes that have different  \textit{spatial and/or temporal extents}. Two agents are not in the same \textit{level} iff they cannot  \textit{interact directly}, \textit{i.e.}, with a single interaction function. It should follow from the previous definitions that it exists multi-scale models that are mono-level and conversely.} agent-based modeling: the conception of a meta-model allowing a non ambiguous characterization of a multi-level model at the conceptual level~\cite{Maus:2008,Morvan:2011,Picault:2011,Uhrmacher:2007}, the introduction of a dynamic level of detail~\cite{Navarro:2011,Soyez:2011}, the detection and reification of emergent phenomena~\cite{Caillou:2012,Chen:2010,David:2009,Moncion:2010,Prevost:2009,Vo:2012} and the representation of aggregated entities~\cite{Parunak:2012}. This paper focuses on the first one, with respect to the Influence Reaction (IR) principle, shortly, action as a two step process: (1) agents produce "influences", \textit{i.e.}, individual decisions, according to their internal state and perceptions (2) the system "reacts", \textit{i.e.}, computes the consequences of influences, according to the state of the world~\cite{Ferber:1996}. This model has been extended in several ways, and notably for multi-agent based-simulation (MABS), by adding an explicit representation of time~\cite{Michel:2007}. An IR-based meta-model, IRM4MLS  (an Influence Reaction Model for Multi-Level Simulation), and its Java implementation are introduced in the section~\ref{irm4mls}.
Using, and then extending a simple example: the Conway's game of life, two toy-models of increasing complexity are presented in the section~\ref{mlgameoflife}, illustrating forms of top-down control in (multi-level) MABS. Results are discussed in the section~\ref{conclusion}. 

\section{IRM4MLS: an Influence Reaction Model for Multi-Level Simulation}
\label{irm4mls}

In this section, IRM4MLS, an Influence Reaction Model for Multi-Level Simulation, is introduced~\cite{Morvan:2011}. It extends IRM\small{4}S  (an Influence Reaction Model for Simulation) in order to deal with multi-level models~\cite{Michel:2007}. From a technical perspective,  levels can be viewed as interacting IRM4S simulations\footnote{Therefore, each level has a microscopic side: the agent behaviors, and a macroscopic side: the reaction function. This aspect can also be found in holonic multi-agent systems~\cite{Cossentino:2010}.}. 

A multi-level model is defined by a set of levels $L$ and a specification of the relations between levels. Two types of relations are specified: an influence relation (agents in a level $l$ are able to produce influences in a level $l' \neq l$) and a perception relation (agents in a level $l$ are able to perceive the dynamic state of a level $l' \neq l$), represented by directed graphs denoted respectively $<L, E_I>$ and $<L, E_P>$, where $E_{I}$  and $E_{P}$ are two sets of edges, \textit{i.e.}, ordered pairs of elements of $L$. Influence and perception relations in a level are systematic and thus not specified in $E_I$ and $E_P$ (cf. eq.~\ref{I-} and \ref{I+}). 

\begin{figure}[hb]
	\begin{center}
		\tikzstyle{blocdebase}= [draw, text centered,minimum height=1cm, minimum width=2cm, rounded corners]

		\begin{tikzpicture}[-,>=stealth',shorten >=1pt,auto,semithick]
   			\node[blocdebase] (agent)  {agent};
   			\node[blocdebase,node distance=\textwidth/5] (environnement)[below left 
of=agent] {environment};
			\node[blocdebase,node distance=\textwidth/5] (niveau)[below right 
of=agent] {level};
    			    
    			\draw (environnement.north) |- (agent.west) ;
			\draw  (niveau.north) |- (agent.east);
			\draw (environnement) -- (niveau);
			
			\node [above right] at (niveau.north) {$0..n$}; 
			\node [above left] at (niveau.west) {$1..n$}; 
			\node [above left] at (agent.west) {$0..n$}; 
			\node [above right] at (agent.east) {$0..n$}; 
			\node [above right] at (environnement.north) {$0..n$}; 
			\node [above right] at (environnement.east) {$1$}; 
						
		\end{tikzpicture}
		\caption{Main concepts of IRM4MLS (as a simplified class diagram)}
		\label{conceptIRM4MLS}
	\end{center}
\end{figure}
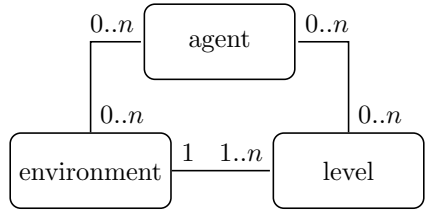

The in and out neighborhood in  $<L, E_I>$ (respectively $<L, E_P>$)  are denoted $N_{I}^-$ and $N_{I}^+$ (resp. $N_{P}^-$ and $N_{P}^+$) and are defined as follows:
\begin{equation}
\forall l \in L, N_{I}^-(l) \mbox{ (resp. } N_{P}^-(l)\mbox{) } =  \{ l \} \cup \{l' \in L: l'l \in E_{I} \mbox{ (resp. } E_{P}\mbox{)} \},
\label{I-} 
\end{equation}
\begin{equation}
\forall l \in L, N_{I}^+(l) \mbox{ (resp. } N_{P}^+(l)\mbox{) } =  \{ l \} \cup \{l' \in L: ll' \in E_{I} \mbox{ (resp. } E_{P}\mbox{)} \},
\label{I+} 
\end{equation}
The set of agents in the system at time $t$ is denoted $A(t)$. $\forall l \in L$, the set of agents belonging to $l$ at $t$ is denoted $A_l(t) \subseteq A(t)$. An agent belongs to a level iff a subset of its physical state $\phi_a$ belongs to the state of the level.
Thus, an agent belongs to zero, one, or more levels. An environment models the \textit{natural dynamics} of level properties and can be shared by different levels (fig. \ref{conceptIRM4MLS}).

The dynamic state of a level $l \in L$ at time $t$, denoted  $\delta^l(t) \in \Delta^l$, is a tuple  $< \sigma^l(t), \gamma^l(t) >$, where $\sigma^l(t)  \in \Sigma^l$ and  $ \gamma^l(t) \in  \Gamma^l$ are the sets of environmental properties and influences of $l$. The behavior of an agent $a \in A_l$  is defined as $Decision_{a}^l \circ Memorization_{a} \circ Perception_{a}^l$, with
\begin{equation}
	\label{perception}
	Perception_{a}^l: \prod_{l_P \in N_P^+(l)}  \Delta^{l_P}  \mapsto   \prod_{l_P \in N_P^+(l)} P_{a}^{l_P},
\end{equation}
\begin{equation}
	Memorization_{a}: \prod_{l \in L | a \in A_l} \prod_{l_P \in N_P^+(l)} P_{a}^{l_P} \times S_{a} \mapsto S_{a}, 
\end{equation}
\begin{equation}
	Decision_{a}^l: S_{a} \mapsto \prod_{l_I \in N_I^+(l)} \Gamma^{l_I}{'}.
\end{equation}
There is no memorization function specific to a level to preserve the coherence of the internal state of the agents. The environment $\omega$ of a level $l$ produces influences through a function: 
\begin{equation}
	Natural_\omega^l: \Delta^{l} \mapsto \prod_{l_I \in N_I^+(l)} \Gamma^{l_I}{'}.
\end{equation}
The reaction function computes next level state and time advance:
 \begin{equation}
	Reaction^l: \Sigma^l \times \Gamma^l{}' \mapsto \Delta^l \times  \mathbb{T}^l.
\end{equation}

The time representation is inspired by DEVS (Discrete EVent System specification)~\cite{Zeigler:2000}.  $\mathbb{T} =  \bigcup_{l \in L} \{ \mathbb{T}^l  \}$ denotes the time vector of the simulation, such as $\forall l \in L, \mathbb{T}^l =<t^l, dt^l>$, where $t^l$ represents when the current event (or step, depending on the simulation model) time and  $dt^l$ its lifespan. The final simulation time is denoted $t_f$. The algorithm~\ref{complexalgo} ensures the scheduling of these different functions with respect to temporal constraints of perception and memorization, influence production and reaction~\cite{Morvan:2011}.

 \begin{algorithm}[h]
	\KwIn{$<L, E_I, E_P >, A, \delta,  \mathbb{T}, t_f$}
	\KwOut{$\delta(t_f)$}
	\While{$\exists t^l \leq t_f$}{
	   \ForEach{$a \in A(t)$}{
	      \ForEach{$ l \in L : a \in A_l(t) \wedge \forall l_P \in N_P^+(l),  t^l \geq t^{l_P}$}{
	        $p_a(t^l) = Perception_a^l(<\delta^{l_P}(t^{l_P}): l_P \in N_P^+(l) >)$\;
	     }
	     $s_a(t^l + dt^l) = Memorization_a(p_a)$\;  
	   }
	   \ForEach{$ l \in L : \forall l_I \in N_I^+(l), t^l \leq t^{l_I}  \vee t^l + dt^l <  t^{l_I} + dt^{l_I}$}{
	   	    \ForEach{$l_I \in N_I^+(l) : t^l \leq  t^{l_I} \wedge  t^l +dt^l > t^{l_I}$}{
		           $\gamma_\omega^{l_I}{}'(t^{l_I}) = Natural_\omega^l(\delta^l(t^l))$ \;
	                    \ForEach{$a \in A_l(t)$}{
	                        $\gamma_a^{l_I}{}'(t^{l_I}) = Decision_a^l(s_a(t^l + dt^l))$\;
	                    }
		    }
	   }
	   \ForEach{$ l \in L : t^l +dt^l  \in min(t + dt)$}{
	          $\gamma^l{'}(t) = \{\gamma^l(t)   \bigcup_{l_I \in N_I^-(l)} \gamma^{l_I}_\omega{'}(t) \bigcup_{a \in A_{l_I}} \gamma_a^{l_I}{'}(t) \}$\;
	          <$\delta^l( t^l +dt^l), \mathbb{T}^l> = Reaction^l(\sigma^l(t^l),  \gamma^l{}'(t^l))$\;
	   }
	}
	\caption{simulation model of IRM4MLS}
	\label{complexalgo}
\end{algorithm}

The implementation of IRM4MLS 
is based on the idea of \textit{micro kernel}, taken from MadKit\footnote{\url{http://www.madkit.org}}~\cite{Gutknecht:2000}. Thus, in this approach, a \textit{technical} agent, \textit{e.g.}, an observer or a message broker, would be scheduled  with respect to the IR principle (cf. algo.~\ref{complexalgo}), the concept of level ensuring a clear separation between system and simulation agents.

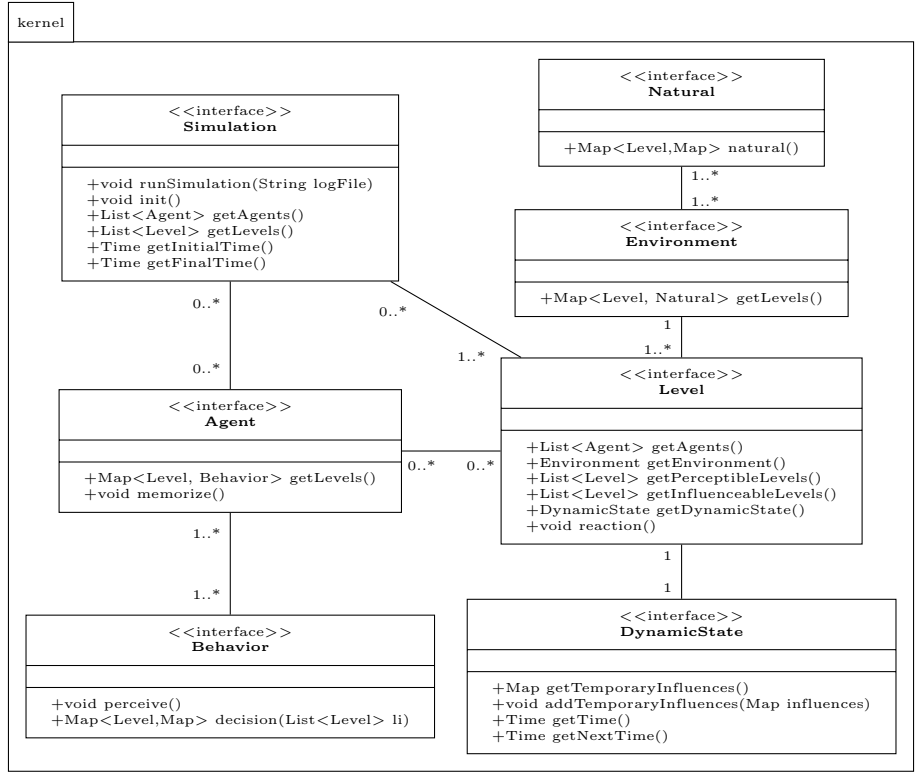
\begin{figure}[h]
	\begin{center}
		\input{figures/api}
		\caption{Java API of IRM4MLS}
		\label{api}
	\end{center}
\end{figure}

The API is minimal (seven \textit{high-level abstractions}) and specifies only the methods needed to schedule a model (fig.~\ref{api}). Most methods are generic and then are implemented at an abstract level. Basically, to implement a model one only has to override  perception, memorization, influence production, reaction and initialization functions.

Agent and behaviors (such as environment and natural dynamics) are represented by different entities to clearly distinguish the \textit{core side} of an agent, its state and memorization function, and the \textit{level sides} (perception and influence production functions) that can change according to simulations.

\section{IRM4MLS in practice: multi-level games of life}
\label{mlgameoflife}

\subsection{Introducing a macroscopic parameter (top-down control)}
\label{gameoflife}

In this section, a toy IRM4S (or $1$-level IRM4MLS) model, is presented:  a modified agent-based version of the Conway's game of life (or simply Life). This simple example illustrates where a macroscopic parameter, should (but should not) be introduced in an agent-based model that relies on the influence reaction principle: in the reaction function (but not in the behavioral functions of  agents). Therefore, this parameter has a non-ambiguous semantics that does not depend on the updating scheme of the simulations, even in the case of \textit{strong interaction}\footnote{Strong interaction implies that agents agree on the outcome of the interaction~\cite{Michel:2003}. Thus,  such model should not be simulated with a STRIPS-like meta-model, \textit{i.e.}, that views an action as a change of the state of the system. It would lead to problems of result replication~\cite{Bigbee:2007}, but also of parameter and result interpretation~\cite{Fates:2010,Michel:2003}.}.

Each agent represents a cell that can be dead or alive and that has eight neighbors in a toroidal grid. The set of environmental properties is then: 
\begin{equation}
 \forall t, \sigma(t) = \bigcup_{a \in A}  \{ neighbors(a), alive(a) \}.
\end{equation}

Cells evolve in parallel: the reaction function can then simply be defined as "applying agent influences"~(algo.~\ref{reactionalgo}). If a cell is dead and has three living neighbors or is alive and has two or three living neighbors, it will be alive at the next step; in other cases, it will be dead. Let specify the behavior of the agents:
\begin{enumerate}
\item they perceive the number of living cells in their neighborhood~(algo.~\ref{perceptionalgo}),
\item memorize their internal state, \textit{i.e.}, their next state~(algo.~\ref{memorizationalgo}),
\item and then, decide whether or not they will be alive  at the next step according to their internal state~(algo.~\ref{decisionalgo}).
\end{enumerate}
The environment is static: there is no natural dynamics and thus, $Natural_\omega$ returns $\emptyset$.

\begin{figure*}[h]
\begin{minipage}[c]{.46\linewidth}
       \begin{algorithm}[H]
	\KwIn{$\sigma(t), \gamma'(t)$}
	\KwOut{$\delta(t+1)$}
	 \ForEach{$a \in A(t)$}{
		$alive(a) = \gamma'_a(t)$ \;
	}
	\caption{$Reaction$}
	\label{reactionalgo}
     \end{algorithm}
     \vspace{0.7cm}
      \begin{algorithm}[H]
	\KwIn{$\delta(t)$}
	\KwOut{$p_a(t)$}
	$p_a(t) = \displaystyle \sum_{n \in neighbors(a)} alive(n)$ \; 
	\caption{$Perception_a$}
	\label{perceptionalgo}
      \end{algorithm}
\end{minipage} \hfill
\begin{minipage}[c]{.46\linewidth}
    \begin{algorithm}[H]
	\KwIn{$p_a(t), s_a(t)$}
	\KwOut{$s_a(t+dt)$}
	\eIf{$alive(a) \wedge  p_a(t) \in \{2, 3\} \vee$\\
		$ \neg alive(a) \wedge  p_a(t) \in \{3\}$
	} {
	   $s_a(t+dt) = 1$ \;
	}
	{
	  $s_a(t+dt) = 0$ \;
	}
	\caption{$Memorization_a$}
	\label{memorizationalgo}
     \end{algorithm}  
     \begin{algorithm}[H]
	\KwIn{$s_a(t+dt)$}
	\KwOut{$\gamma'_a(t)$}
	$\gamma'_a(t) = s_a(t+dt)$ \;
	\caption{$Decision_a$}
	\label{decisionalgo}
     \end{algorithm}
  
\end{minipage}
\end{figure*}


\begin{figure}[t]
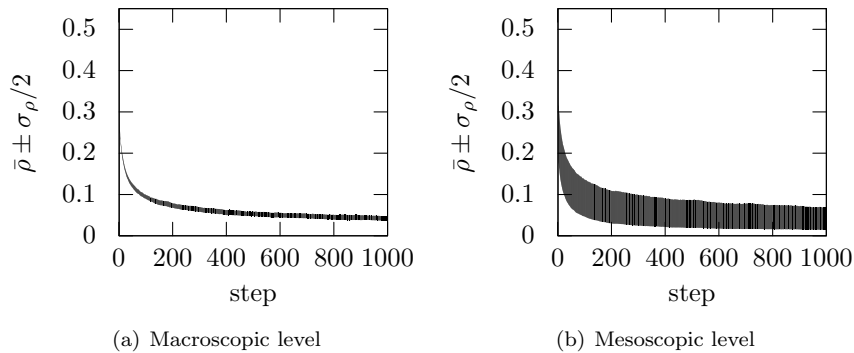

	\begin{center}
		\subfigure[Macroscopic level]{\input{figures/longRunPopulation}\label{longRunPopulation}}
		\subfigure[Mesoscopic level]{\input{figures/longRunVariability}\label{longRunCluster}}
		\caption{Dynamics of the density of living cells,  $\rho$, starting from a random grid with 100 replications (a) Macroscopic dynamics: expected density and variability of living cells in the whole grid. (b) Mesoscopic  dynamics: expected density and variability of living cells in $10\times10$ cell clusters.}
	\end{center}
\end{figure}

One cruel and ironic aspect of Life is that a cell has generally little chance to remain alive in the long run (fig. \ref{longRunPopulation}). Moreover, what you get most of the time, is a board composed of small still lifes and $1$-period oscillators. This behavior is predictable knowing the $\lambda$ parameter of the game of life\footnote{$\lambda_{life} = 0.2734375$. For $\lambda \approx 0.25$, \textit{"structures of period 1 appear. Thus, there are now three different possible outcomes for the ultimate dynamics of the system, depending on the initial state. The dynamics may reach a homogeneous fixed point consisting entirely of state $S_q$, or it may reach a heterogeneous fixed point consisting mostly of cells in state $S_q$ with a sprinkling of cells stuck in one of the other states, or it may settle down to periodic behavior"}~\cite[p. 17]{Langton:1990}.}. $\lambda$ is a complexity measure of cellular automata introduced by~\cite{Langton:1990} that depends on the number of cell states $K$, the neighborhood $N$ and the number $n$ of transitions to a quiescent state $S_q$ in the transition function such as
\begin{equation}
	\lambda = 1 - \frac{n}{K^N},
\end{equation}
with $K = 2$, $N = 9$ and $n = \tbinom 82 + 2 \cdot \sum_{i = 0,1,4-8} \tbinom 8i$ for Life.

\begin{figure}[ht]
	\begin{center}
		\subfigure[Convergence conditions]{\input{figures/stilllife}}
		\subfigure[Simulation example]{\raisebox{0.5cm}{\includegraphics[]{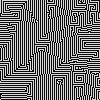}}}
		\caption{(a) Mean number of steps needed to converge to a steady state starting from a random grid ($100$ replications), simulations are stopped after $2 \cdot 10^4$ steps (simulations converge in the dark area). (b) Example of still life found for $\lambda^+ = 0.5$.}
		\label{stilllife}
	\end{center}
\end{figure}

To improve $\lambda_{life}$ to a value $\lambda^+$, one needs to change the rules. In this example, $\lambda^{+}$ is regarded as a macroscopic parameter, explicitly introduced in the model and independent from cell behaviors\footnote{An other macroscopic parameter, the asynchrony, has been previously introduced in Life is such way~\cite{Fates:2010a}.}: influences of dying cells are not taken into account by the reaction function with a probability $p$ such as
\begin{equation}
	n = \tbinom 82 + (2 - p) \cdot \displaystyle \sum_{i = 0,1,4-8} \tbinom 8i.
\end{equation}
Thus, their is a simple linear relation between $p$ and $\lambda^{+}$: $p = (\lambda^{+} - \lambda_{life}) / 0.3359$. For $\lambda^{+} \in [0.48, 0.6]$, the number of dying cells tends to decrease in time and large structures of vertical or horizontal rows eventually emerge, shaped by moving groups of switching state cells that seem to work at their boundaries, and eventually vanish when the board becomes a dense still life of density $\approx 0.5$ (fig.~\ref{stilllife}).

\subsection{Top-down feedback control}

In the previous example, while the macroscopic parameter $\lambda^{+}$ has an influence on agents,  it is not related to the state of the system and therefore, there is no need to observe it. The reaction function can then be viewed as an \textit{open-loop} controller. In the next example, a top-down feedback control is introduced.

The goal of the multi-scale model presented in this section is to keep Life boards at the desired density ($\rho^+$), by controlling the proportion of dying cells at the mesoscopic level, to account for the \textit{natural} variability of density between regions of the grid (fig.~\ref{longRunCluster}).  Moreover, the control should affect \textit{as less as possible} simulations at the microscopic level and, to keep it simple, should be tuned by a single linear parameter such as $\lambda^{+}$ in the previous model.  

Two levels are considered: the cell (or microscopic) level, $l_m$ and the cell region (or mesoscopic) level, $l_M$. At the mesoscopic level, the model properties are the expected density and the cells in each region:
\begin{equation}
 \forall t, \sigma^{l_M}(t) = \{\rho^+ \}  \bigcup_{a^{l_M} \in A^{l_M}}  \{cells(a^{l_M}) \}.
\end{equation}
The cells behave according to the game of life rules  (algo.~\ref{perceptionalgo},~\ref{memorizationalgo} and~\ref{decisionalgo}). However, $Reaction^{l_m}$ depends on mesoscopic influences (algorithm~\ref{reactionalgo2}). $E_{P}$ and $E_{I}$ are equal to $\{ l_Ml_m \}$. Mesoscopic  agents have a proportional control behavior. They
\begin{enumerate}
\item perceive the density of living cells in a region,
\item memorize their internal state, \textit{i.e.}, the difference $\epsilon$ between expected and actual densities,
\item and then decide the influence sent to agents of ${l_m}$: 
\begin{equation}
\forall a^{l_M} \in A^{l_M} \forall a^{l_m} \in  cells(a^{l_M}), command(a^{l_m}) =  k_P \cdot \epsilon. 
\end{equation}
\end{enumerate}

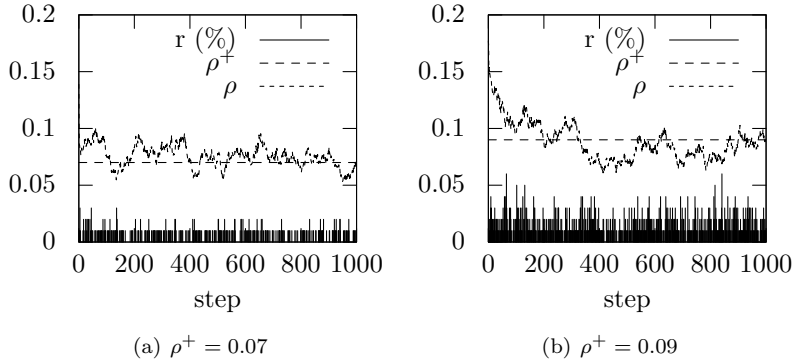
\begin{figure}[t]
	\begin{center}
		\subfigure[$\rho^+ = 0.07$]{\input{figures/ml007}}
		\subfigure[$\rho^+ = 0.09$]{\input{figures/ml009}}
		\caption{Simulation examples with $k_P = 10 \cdot \rho^+$ and initial density $\rho(t_0) = 2 \cdot \rho^+$. $r$ represents the rate of microscopic influences not taken into account  (in \%) by $Reaction^{l_m}$.}
		\label{mlfig}
	\end{center}
\end{figure}

The $k_P$ parameter has to be carefully tuned to run realistic simulations:  too small simulations do not achieve the desired solution ($\bar{\rho} =  \rho^+$), too big the board density tends to oscillate around  $\rho^+$ and  the number of micro influences not taken into account by $Reaction^{l_m}$ becomes too important. However, for appropriate $k_P$ values, this simple linear controller achieves good results and allows to find a good compromise between conflicting micro and meso knowledge~(fig.~\ref{mlfig}).

\begin{algorithm}[t]
	\KwIn{$\sigma^{l_m}(t), \gamma^{l_m}{'}(t)$}
	\KwOut{$\delta^{l_m}(t+1)$}
	 \ForEach{$a^{l_m} \in a^{l_m}$}{
	 	$rand \in [0, 1[$ from pseudorandom uniform distribution \;
		\eIf{$command(a^{l_m}) > rand \wedge alive(a^{l_m}) \wedge \neg \gamma{'}_{a^{l_m}}(t^{l_m})$}{
			$alive(a^{l_m}) = \top$ \;
		}{
			$alive(a^{l_m}) = \gamma{'}_{a^{l_m}}(t^{l_m})$ \;
		}
	 }
	\caption{$Reaction^{l_m}$}
	\label{reactionalgo2}
\end{algorithm}

%

\section{Conclusion}
\label{conclusion}

\subsection{Discussion}

An important issue of multi-level agent-based modeling, only briefly discussed here, is to define the adequate methodology. Indeed, the \textit{traditional} MABS methodology is purely bottom-up: microscopic knowledge is used to construct models while macroscopic  knowledge is used to validate models~\cite{Drogoul:2003a}. Thus, it seems irrelevant in a multi-level context. Three general conclusions can be drawn from the previous examples:
\begin{itemize}
\item a parameter should be introduced at its observation level in the model; therefore, each observed level should be explicitly represented in the model,
\item an intuitive way to model an external control on a level $l$ is to modify the reaction function of $l$, \textit{i.e.}, to modify the way influences of agents of $l$ are taken into account (but not agent behavioral functions), according to external influences; an external feedback control implies both observation and  influence relations: $N_{I}^-(l) = N_{P}^-(l)$,
\item such a controller can be viewed as a technical tool that aims to find a compromise between conflicting knowledge from the different studied levels and achieve realistic simulations from different studied points of view~\cite{Li:2005}.
\end{itemize}

Pattern oriented modeling (POM) consists in \textit{"the multi-criteria design, selection and calibration of models of complex systems"}~\cite{Grimm:2012}. Many aspects of this methodology, developed in the context of ecology, seem particularly relevant for multi-level agent-based models since "patterns"  are generally observed at different levels of organization in complex systems. However, the problem is far from being solved. For instance, the introduction of a dynamic level of detail raises several questions regarding, \textit{e.g.}, the validation of the model or the representation of \textit{composite} agents~\cite{Navarro:2011,Parunak:2012}. A case study of three real world multi-level agent-based models reveals other interesting methodological issues~\cite{Gil-Quijano:2012}. 

\subsection{Related works}

At least three works could be related to this one:
\begin{itemize}
\item ML-DEVS is an extension of DEVS that allows the simulation of multi-level models (and not only coupled models in which the behavior of a model is determined by the behaviors of its sub-models)~\cite{Uhrmacher:2007}. Two types of relation between levels are defined: information propagation and event activation which are quite similar to those defined in IRM4MLS. However, ML-DEVS supports only pure hierarchies of models, \textit{i.e.}, \textit{interaction graphs} are viewed as \textit{trees}~\cite{Maus:2008}. DEVS, as a generic event-based simulation framework, has also  been extended to support agent-based models~\cite{Muller:2009}. A major design difference between IR and DEVS based approaches is the technical orientation of the latter leading to an important gap between conceptual and computational models.
\item  PADAWAN (Pattern for Accurate Design of Agent Worlds in Agent Nests) is a multi-scale agent-based meta-model based on a compact matricial representation of interactions: IODA (Interaction-Oriented Design of Agent simulations)~\cite{Kubera:2008,Picault:2011}. Moreover, authors analyze the structure of what is called here interaction graphs in multi-scale models (a relevant issue for IRM4MLS as well), and conclude they should be viewed as \textit{upper semilattices} and not simply \textit{trees} as suggested elsewhere. A major design difference between IR and IODA based approaches is that the latter constraints the definition of interactions, leading to a simple but restrictive simulation framework.
\item GAMA\footnote{\url{http://code.google.com/p/gama-platform/}} is a MABS platform with a dedicated modeling language, GAML, that offers multi-level capabilities~\cite{Taillandier:2012}. Moreover, it includes a framework (a set of predefined GAML commands) to \textit{agentify} emerging structures~\cite{Vo:2012}. It is certainly the most advanced platform, from an end-user point of view, that integrates  a multi-level approach.
\end{itemize}

\subsection{Perspectives}

The main perspectives of this work concern the implementation of existing works with IRM4MLS: 
\begin{itemize}
\item the concept of \textit{PolyAgent}~\cite{Parunak:2011,Parunak:2007},
\item multi-level organizational models widely used in engineering sciences such as holonic multi-agent systems (cf. footnote~2), system of systems and heterarchical control~\cite{Cossentino:2010,Maier:1998,Morvan:2012a},
\item multi-scale tools: generic scaling operators and emergence detection and reification algorithms~\cite{Chen:2010,Navarro:2011}. 
\end{itemize}

Moreover, the first model presented in this paper could be used to explore the relations between computational capabilities of a cellular automaton ($\lambda^{+}$), noise (a function of $\lambda^{+} - \lambda$) and entropy. Moreover, finding the conditions under which cells arrange themselves in a steady state could be an interesting way to solve heuristically large instances of the \textit{maximum density still life problem}~\cite{Elkies:1998}.

The second model illustrates a form of simple proportional top-down feedback control. Such approach could be generalized to model more complex of cross-level feedback control, using integrations and derivates of observed variables.


%

\bibliographystyle{splncs03.bst}
\bibliography{../../Biblio}

\end{document}

%% file: figures/api.tex
\tikzumlset{font=\tiny}
\begin{tikzpicture}

\begin{umlpackage}[fill=white]{kernel}
	\umlclass [type=interface,x=-3,y=0, fill=white] {Agent}{}{ 
	+Map<Level, Behavior> getLevels() \\
	+void memorize()
	}
	\umlclass [type=interface,x=-3,y=-3, fill=white] {Behavior}{
	}{ 
	+void perceive()\\
	+Map<Level,Map> decision(List<Level> li)
	}
	
	\umlclass [type=interface,x=3,y=0,fill=white] {Level}{
	}{ 
	+List<Agent> getAgents()\\
	+Environment getEnvironment() \\
	+List<Level> getPerceptibleLevels()\\
	+List<Level> getInfluenceableLevels()\\
	+DynamicState getDynamicState()\\
	+void reaction()
	}
	
	\umlclass [type=interface,x=3,y=-3,fill=white] {DynamicState}{
	}{ 
	+Map getTemporaryInfluences()\\
	+void addTemporaryInfluences(Map influences)\\
	+Time getTime()\\
	+Time getNextTime()
	}
	
	\umlclass [type=interface,x=3,y=2.5,fill=white] {Environment}{
	}{ 
	+Map<Level, Natural> getLevels()
	}
	
	\umlclass [type=interface,x=3,y=4.5,fill=white] {Natural}{
	}{ 
	+Map<Level,Map> natural()
	}
	
	\umlclass [type=interface,x=-3,y=3.5,fill=white] {Simulation}{
	}{ 
	+void runSimulation(String logFile)\\
	+void init()\\
	+List<Agent> getAgents()\\
	+List<Level> getLevels()\\
	+Time getInitialTime()\\
	+Time getFinalTime()
	}
	\umlassoc[mult1=1, mult2=1]{Level}{DynamicState}
	\umlassoc[mult1=1..*, mult2=1..*]{Environment}{Natural}
	\umlassoc[mult1=1, mult2=1..*]{Environment}{Level}
	\umlassoc[mult1=0..*, mult2=0..*]{Agent}{Level}
	\umlassoc[mult1=1..*, mult2=1..*]{Agent}{Behavior}
	\umlassoc[mult1=0..*, mult2=0..*]{Simulation}{Agent}
	\umlassoc[mult1=0..*, mult2=1..*]{Simulation}{Level}

\end{umlpackage}
\end{tikzpicture}

%% file: figures/stilllife.tex
\begin{tikzpicture}[gnuplot]
\gpmonochromelines
\gpcolor{gp lt color border}
\gpsetlinetype{gp lt border}
\gpsetlinewidth{1.00}
\draw[gp path] (1.872,0.985)--(2.052,0.985);
\draw[gp path] (5.072,0.985)--(4.892,0.985);
\node[gp node right] at (1.688,0.985) { 8000};
\draw[gp path] (1.872,1.489)--(2.052,1.489);
\draw[gp path] (5.072,1.489)--(4.892,1.489);
\node[gp node right] at (1.688,1.489) { 10000};
\draw[gp path] (1.872,1.992)--(2.052,1.992);
\draw[gp path] (5.072,1.992)--(4.892,1.992);
\node[gp node right] at (1.688,1.992) { 12000};
\draw[gp path] (1.872,2.496)--(2.052,2.496);
\draw[gp path] (5.072,2.496)--(4.892,2.496);
\node[gp node right] at (1.688,2.496) { 14000};
\draw[gp path] (1.872,3.000)--(2.052,3.000);
\draw[gp path] (5.072,3.000)--(4.892,3.000);
\node[gp node right] at (1.688,3.000) { 16000};
\draw[gp path] (1.872,3.503)--(2.052,3.503);
\draw[gp path] (5.072,3.503)--(4.892,3.503);
\node[gp node right] at (1.688,3.503) { 18000};
\draw[gp path] (1.872,4.007)--(2.052,4.007);
\draw[gp path] (5.072,4.007)--(4.892,4.007);
\node[gp node right] at (1.688,4.007) { 20000};
\draw[gp path] (1.872,0.985)--(1.872,1.165);
\draw[gp path] (1.872,4.007)--(1.872,3.827);
\node[gp node center] at (1.872,0.677) { 0.45};
\draw[gp path] (2.672,0.985)--(2.672,1.165);
\draw[gp path] (2.672,4.007)--(2.672,3.827);
\node[gp node center] at (2.672,0.677) { 0.5};
\draw[gp path] (3.472,0.985)--(3.472,1.165);
\draw[gp path] (3.472,4.007)--(3.472,3.827);
\node[gp node center] at (3.472,0.677) { 0.55};
\draw[gp path] (4.272,0.985)--(4.272,1.165);
\draw[gp path] (4.272,4.007)--(4.272,3.827);
\node[gp node center] at (4.272,0.677) { 0.6};
\draw[gp path] (5.072,0.985)--(5.072,1.165);
\draw[gp path] (5.072,4.007)--(5.072,3.827);
\node[gp node center] at (5.072,0.677) { 0.65};
\draw[gp path] (1.872,4.007)--(1.872,0.985)--(5.072,0.985)--(5.072,4.007)--cycle;
\node[gp node center,rotate=-270] at (0.246,2.496) {number of steps};
\node[gp node center] at (3.472,0.215) {$\lambda^+$};
\gpfill{color=gp lt color 0} (1.872,4.007)--(1.872,4.007)--(2.032,4.007)--(2.192,4.007)%
    --(2.352,4.007)--(2.512,1.095)--(2.672,1.658)--(2.832,1.177)--(2.992,1.489)%
    --(3.152,1.816)--(3.312,2.283)--(3.472,2.643)--(3.632,2.843)--(3.792,3.052)%
    --(3.952,3.528)--(4.112,3.811)--(4.272,4.007)--(4.432,4.007)--(4.592,4.007)%
    --(4.752,4.007)--(4.912,4.007)--(5.072,4.007)--cycle;
\gpcolor{gp lt color 0}
\gpsetlinetype{gp lt plot 0}
\draw[gp path] (1.872,4.007)--(2.032,4.007)--(2.192,4.007)--(2.352,4.007)--(2.512,1.095)%
  --(2.672,1.658)--(2.832,1.177)--(2.992,1.489)--(3.152,1.816)--(3.312,2.283)--(3.472,2.643)%
  --(3.632,2.843)--(3.792,3.052)--(3.952,3.528)--(4.112,3.811)--(4.272,4.007)--(4.432,4.007)%
  --(4.592,4.007)--(4.752,4.007)--(4.912,4.007)--(5.072,4.007);
\gpcolor{gp lt color border}
\gpsetlinetype{gp lt border}
\draw[gp path] (1.872,4.007)--(1.872,0.985)--(5.072,0.985)--(5.072,4.007)--cycle;
\gpdefrectangularnode{gp plot 1}{\pgfpoint{1.872cm}{0.985cm}}{\pgfpoint{5.072cm}{4.007cm}}
\end{tikzpicture}

%% file: figures/ml007.tex
\begin{tikzpicture}[gnuplot]
\gpmonochromelines
\gpcolor{gp lt color border}
\gpsetlinetype{gp lt border}
\gpsetlinewidth{1.00}
\draw[gp path] (1.380,0.985)--(1.560,0.985);
\draw[gp path] (5.072,0.985)--(4.892,0.985);
\node[gp node right] at (1.196,0.985) { 0};
\draw[gp path] (1.380,1.741)--(1.560,1.741);
\draw[gp path] (5.072,1.741)--(4.892,1.741);
\node[gp node right] at (1.196,1.741) { 0.05};
\draw[gp path] (1.380,2.496)--(1.560,2.496);
\draw[gp path] (5.072,2.496)--(4.892,2.496);
\node[gp node right] at (1.196,2.496) { 0.1};
\draw[gp path] (1.380,3.252)--(1.560,3.252);
\draw[gp path] (5.072,3.252)--(4.892,3.252);
\node[gp node right] at (1.196,3.252) { 0.15};
\draw[gp path] (1.380,4.007)--(1.560,4.007);
\draw[gp path] (5.072,4.007)--(4.892,4.007);
\node[gp node right] at (1.196,4.007) { 0.2};
\draw[gp path] (1.380,0.985)--(1.380,1.165);
\draw[gp path] (1.380,4.007)--(1.380,3.827);
\node[gp node center] at (1.380,0.677) { 0};
\draw[gp path] (2.118,0.985)--(2.118,1.165);
\draw[gp path] (2.118,4.007)--(2.118,3.827);
\node[gp node center] at (2.118,0.677) { 200};
\draw[gp path] (2.857,0.985)--(2.857,1.165);
\draw[gp path] (2.857,4.007)--(2.857,3.827);
\node[gp node center] at (2.857,0.677) { 400};
\draw[gp path] (3.595,0.985)--(3.595,1.165);
\draw[gp path] (3.595,4.007)--(3.595,3.827);
\node[gp node center] at (3.595,0.677) { 600};
\draw[gp path] (4.334,0.985)--(4.334,1.165);
\draw[gp path] (4.334,4.007)--(4.334,3.827);
\node[gp node center] at (4.334,0.677) { 800};
\draw[gp path] (5.072,0.985)--(5.072,1.165);
\draw[gp path] (5.072,4.007)--(5.072,3.827);
\node[gp node center] at (5.072,0.677) { 1000};
\draw[gp path] (1.380,4.007)--(1.380,0.985)--(5.072,0.985)--(5.072,4.007)--cycle;
\node[gp node center] at (3.226,0.215) {step};
\node[gp node right] at (3.604,3.673) {r (\%)};
\gpcolor{gp lt color 0}
\gpsetlinetype{gp lt plot 0}
\draw[gp path] (3.788,3.673)--(4.704,3.673);
\draw[gp path] (1.387,0.985)--(1.387,1.136);
\draw[gp path] (1.391,0.985)--(1.391,1.136);
\draw[gp path] (1.395,0.985)--(1.395,1.438);
\draw[gp path] (1.402,0.985)--(1.402,1.136);
\draw[gp path] (1.435,0.985)--(1.435,1.136);
\draw[gp path] (1.443,0.985)--(1.443,1.287);
\draw[gp path] (1.461,0.985)--(1.461,1.136);
\draw[gp path] (1.472,0.985)--(1.472,1.287);
\draw[gp path] (1.476,0.985)--(1.476,1.287);
\draw[gp path] (1.480,0.985)--(1.480,1.136);
\draw[gp path] (1.498,0.985)--(1.498,1.287);
\draw[gp path] (1.502,0.985)--(1.502,1.136);
\draw[gp path] (1.506,0.985)--(1.506,1.136);
\draw[gp path] (1.509,0.985)--(1.509,1.287);
\draw[gp path] (1.528,0.985)--(1.528,1.287);
\draw[gp path] (1.535,0.985)--(1.535,1.136);
\draw[gp path] (1.546,0.985)--(1.546,1.438);
\draw[gp path] (1.550,0.985)--(1.550,1.136);
\draw[gp path] (1.557,0.985)--(1.557,1.136);
\draw[gp path] (1.565,0.985)--(1.565,1.136);
\draw[gp path] (1.594,0.985)--(1.594,1.136);
\draw[gp path] (1.598,0.985)--(1.598,1.136);
\draw[gp path] (1.609,0.985)--(1.609,1.136);
\draw[gp path] (1.627,0.985)--(1.627,1.136);
\draw[gp path] (1.642,0.985)--(1.642,1.136);
\draw[gp path] (1.646,0.985)--(1.646,1.136);
\draw[gp path] (1.657,0.985)--(1.657,1.136);
\draw[gp path] (1.668,0.985)--(1.668,1.136);
\draw[gp path] (1.675,0.985)--(1.675,1.136);
\draw[gp path] (1.683,0.985)--(1.683,1.136);
\draw[gp path] (1.686,0.985)--(1.686,1.136);
\draw[gp path] (1.690,0.985)--(1.690,1.136);
\draw[gp path] (1.701,0.985)--(1.701,1.287);
\draw[gp path] (1.712,0.985)--(1.712,1.136);
\draw[gp path] (1.723,0.985)--(1.723,1.136);
\draw[gp path] (1.727,0.985)--(1.727,1.287);
\draw[gp path] (1.731,0.985)--(1.731,1.136);
\draw[gp path] (1.734,0.985)--(1.734,1.136);
\draw[gp path] (1.738,0.985)--(1.738,1.136);
\draw[gp path] (1.760,0.985)--(1.760,1.136);
\draw[gp path] (1.768,0.985)--(1.768,1.136);
\draw[gp path] (1.775,0.985)--(1.775,1.136);
\draw[gp path] (1.790,0.985)--(1.790,1.136);
\draw[gp path] (1.794,0.985)--(1.794,1.287);
\draw[gp path] (1.808,0.985)--(1.808,1.136);
\draw[gp path] (1.816,0.985)--(1.816,1.136);
\draw[gp path] (1.819,0.985)--(1.819,1.136);
\draw[gp path] (1.834,0.985)--(1.834,1.136);
\draw[gp path] (1.838,0.985)--(1.838,1.136);
\draw[gp path] (1.849,0.985)--(1.849,1.136);
\draw[gp path] (1.860,0.985)--(1.860,1.136);
\draw[gp path] (1.864,0.985)--(1.864,1.136);
\draw[gp path] (1.871,0.985)--(1.871,1.136);
\draw[gp path] (1.875,0.985)--(1.875,1.136);
\draw[gp path] (1.882,0.985)--(1.882,1.438);
\draw[gp path] (1.889,0.985)--(1.889,1.287);
\draw[gp path] (1.893,0.985)--(1.893,1.136);
\draw[gp path] (1.919,0.985)--(1.919,1.136);
\draw[gp path] (1.930,0.985)--(1.930,1.136);
\draw[gp path] (1.941,0.985)--(1.941,1.136);
\draw[gp path] (1.949,0.985)--(1.949,1.136);
\draw[gp path] (1.956,0.985)--(1.956,1.136);
\draw[gp path] (1.963,0.985)--(1.963,1.136);
\draw[gp path] (1.967,0.985)--(1.967,1.136);
\draw[gp path] (1.971,0.985)--(1.971,1.136);
\draw[gp path] (1.997,0.985)--(1.997,1.136);
\draw[gp path] (2.063,0.985)--(2.063,1.136);
\draw[gp path] (2.081,0.985)--(2.081,1.136);
\draw[gp path] (2.089,0.985)--(2.089,1.136);
\draw[gp path] (2.096,0.985)--(2.096,1.136);
\draw[gp path] (2.133,0.985)--(2.133,1.287);
\draw[gp path] (2.155,0.985)--(2.155,1.136);
\draw[gp path] (2.174,0.985)--(2.174,1.136);
\draw[gp path] (2.177,0.985)--(2.177,1.136);
\draw[gp path] (2.200,0.985)--(2.200,1.287);
\draw[gp path] (2.207,0.985)--(2.207,1.136);
\draw[gp path] (2.218,0.985)--(2.218,1.136);
\draw[gp path] (2.259,0.985)--(2.259,1.136);
\draw[gp path] (2.262,0.985)--(2.262,1.136);
\draw[gp path] (2.270,0.985)--(2.270,1.136);
\draw[gp path] (2.273,0.985)--(2.273,1.136);
\draw[gp path] (2.292,0.985)--(2.292,1.136);
\draw[gp path] (2.299,0.985)--(2.299,1.136);
\draw[gp path] (2.310,0.985)--(2.310,1.287);
\draw[gp path] (2.314,0.985)--(2.314,1.136);
\draw[gp path] (2.318,0.985)--(2.318,1.136);
\draw[gp path] (2.347,0.985)--(2.347,1.136);
\draw[gp path] (2.355,0.985)--(2.355,1.136);
\draw[gp path] (2.373,0.985)--(2.373,1.136);
\draw[gp path] (2.377,0.985)--(2.377,1.136);
\draw[gp path] (2.384,0.985)--(2.384,1.136);
\draw[gp path] (2.388,0.985)--(2.388,1.136);
\draw[gp path] (2.392,0.985)--(2.392,1.136);
\draw[gp path] (2.406,0.985)--(2.406,1.136);
\draw[gp path] (2.417,0.985)--(2.417,1.136);
\draw[gp path] (2.429,0.985)--(2.429,1.136);
\draw[gp path] (2.432,0.985)--(2.432,1.136);
\draw[gp path] (2.440,0.985)--(2.440,1.136);
\draw[gp path] (2.454,0.985)--(2.454,1.136);
\draw[gp path] (2.458,0.985)--(2.458,1.136);
\draw[gp path] (2.469,0.985)--(2.469,1.136);
\draw[gp path] (2.480,0.985)--(2.480,1.136);
\draw[gp path] (2.506,0.985)--(2.506,1.136);
\draw[gp path] (2.513,0.985)--(2.513,1.136);
\draw[gp path] (2.521,0.985)--(2.521,1.136);
\draw[gp path] (2.536,0.985)--(2.536,1.287);
\draw[gp path] (2.547,0.985)--(2.547,1.136);
\draw[gp path] (2.569,0.985)--(2.569,1.136);
\draw[gp path] (2.573,0.985)--(2.573,1.136);
\draw[gp path] (2.584,0.985)--(2.584,1.136);
\draw[gp path] (2.595,0.985)--(2.595,1.136);
\draw[gp path] (2.602,0.985)--(2.602,1.136);
\draw[gp path] (2.606,0.985)--(2.606,1.136);
\draw[gp path] (2.624,0.985)--(2.624,1.136);
\draw[gp path] (2.650,0.985)--(2.650,1.136);
\draw[gp path] (2.657,0.985)--(2.657,1.287);
\draw[gp path] (2.661,0.985)--(2.661,1.136);
\draw[gp path] (2.665,0.985)--(2.665,1.287);
\draw[gp path] (2.669,0.985)--(2.669,1.136);
\draw[gp path] (2.672,0.985)--(2.672,1.136);
\draw[gp path] (2.713,0.985)--(2.713,1.136);
\draw[gp path] (2.728,0.985)--(2.728,1.136);
\draw[gp path] (2.735,0.985)--(2.735,1.136);
\draw[gp path] (2.742,0.985)--(2.742,1.136);
\draw[gp path] (2.753,0.985)--(2.753,1.136);
\draw[gp path] (2.761,0.985)--(2.761,1.136);
\draw[gp path] (2.801,0.985)--(2.801,1.136);
\draw[gp path] (2.816,0.985)--(2.816,1.136);
\draw[gp path] (2.835,0.985)--(2.835,1.136);
\draw[gp path] (2.849,0.985)--(2.849,1.136);
\draw[gp path] (2.894,0.985)--(2.894,1.287);
\draw[gp path] (2.901,0.985)--(2.901,1.136);
\draw[gp path] (2.905,0.985)--(2.905,1.136);
\draw[gp path] (2.908,0.985)--(2.908,1.136);
\draw[gp path] (2.916,0.985)--(2.916,1.287);
\draw[gp path] (2.931,0.985)--(2.931,1.136);
\draw[gp path] (2.945,0.985)--(2.945,1.136);
\draw[gp path] (2.949,0.985)--(2.949,1.136);
\draw[gp path] (2.964,0.985)--(2.964,1.136);
\draw[gp path] (2.968,0.985)--(2.968,1.136);
\draw[gp path] (2.979,0.985)--(2.979,1.136);
\draw[gp path] (2.982,0.985)--(2.982,1.136);
\draw[gp path] (2.990,0.985)--(2.990,1.136);
\draw[gp path] (2.993,0.985)--(2.993,1.136);
\draw[gp path] (3.012,0.985)--(3.012,1.136);
\draw[gp path] (3.056,0.985)--(3.056,1.136);
\draw[gp path] (3.064,0.985)--(3.064,1.136);
\draw[gp path] (3.067,0.985)--(3.067,1.136);
\draw[gp path] (3.075,0.985)--(3.075,1.136);
\draw[gp path] (3.082,0.985)--(3.082,1.136);
\draw[gp path] (3.086,0.985)--(3.086,1.136);
\draw[gp path] (3.108,0.985)--(3.108,1.136);
\draw[gp path] (3.119,0.985)--(3.119,1.136);
\draw[gp path] (3.123,0.985)--(3.123,1.136);
\draw[gp path] (3.126,0.985)--(3.126,1.136);
\draw[gp path] (3.156,0.985)--(3.156,1.136);
\draw[gp path] (3.160,0.985)--(3.160,1.136);
\draw[gp path] (3.182,0.985)--(3.182,1.136);
\draw[gp path] (3.185,0.985)--(3.185,1.136);
\draw[gp path] (3.200,0.985)--(3.200,1.136);
\draw[gp path] (3.204,0.985)--(3.204,1.136);
\draw[gp path] (3.233,0.985)--(3.233,1.287);
\draw[gp path] (3.237,0.985)--(3.237,1.287);
\draw[gp path] (3.244,0.985)--(3.244,1.136);
\draw[gp path] (3.252,0.985)--(3.252,1.136);
\draw[gp path] (3.256,0.985)--(3.256,1.136);
\draw[gp path] (3.259,0.985)--(3.259,1.136);
\draw[gp path] (3.267,0.985)--(3.267,1.136);
\draw[gp path] (3.270,0.985)--(3.270,1.136);
\draw[gp path] (3.278,0.985)--(3.278,1.136);
\draw[gp path] (3.296,0.985)--(3.296,1.136);
\draw[gp path] (3.307,0.985)--(3.307,1.136);
\draw[gp path] (3.311,0.985)--(3.311,1.136);
\draw[gp path] (3.318,0.985)--(3.318,1.287);
\draw[gp path] (3.333,0.985)--(3.333,1.136);
\draw[gp path] (3.344,0.985)--(3.344,1.136);
\draw[gp path] (3.348,0.985)--(3.348,1.136);
\draw[gp path] (3.352,0.985)--(3.352,1.136);
\draw[gp path] (3.363,0.985)--(3.363,1.136);
\draw[gp path] (3.381,0.985)--(3.381,1.136);
\draw[gp path] (3.388,0.985)--(3.388,1.136);
\draw[gp path] (3.414,0.985)--(3.414,1.136);
\draw[gp path] (3.418,0.985)--(3.418,1.136);
\draw[gp path] (3.425,0.985)--(3.425,1.136);
\draw[gp path] (3.440,0.985)--(3.440,1.136);
\draw[gp path] (3.451,0.985)--(3.451,1.287);
\draw[gp path] (3.455,0.985)--(3.455,1.287);
\draw[gp path] (3.459,0.985)--(3.459,1.136);
\draw[gp path] (3.525,0.985)--(3.525,1.136);
\draw[gp path] (3.536,0.985)--(3.536,1.287);
\draw[gp path] (3.540,0.985)--(3.540,1.287);
\draw[gp path] (3.547,0.985)--(3.547,1.136);
\draw[gp path] (3.562,0.985)--(3.562,1.287);
\draw[gp path] (3.580,0.985)--(3.580,1.136);
\draw[gp path] (3.610,0.985)--(3.610,1.136);
\draw[gp path] (3.621,0.985)--(3.621,1.136);
\draw[gp path] (3.632,0.985)--(3.632,1.136);
\draw[gp path] (3.640,0.985)--(3.640,1.136);
\draw[gp path] (3.643,0.985)--(3.643,1.287);
\draw[gp path] (3.654,0.985)--(3.654,1.136);
\draw[gp path] (3.658,0.985)--(3.658,1.136);
\draw[gp path] (3.665,0.985)--(3.665,1.136);
\draw[gp path] (3.684,0.985)--(3.684,1.136);
\draw[gp path] (3.699,0.985)--(3.699,1.136);
\draw[gp path] (3.706,0.985)--(3.706,1.136);
\draw[gp path] (3.713,0.985)--(3.713,1.136);
\draw[gp path] (3.717,0.985)--(3.717,1.136);
\draw[gp path] (3.735,0.985)--(3.735,1.136);
\draw[gp path] (3.747,0.985)--(3.747,1.136);
\draw[gp path] (3.750,0.985)--(3.750,1.287);
\draw[gp path] (3.754,0.985)--(3.754,1.136);
\draw[gp path] (3.765,0.985)--(3.765,1.136);
\draw[gp path] (3.791,0.985)--(3.791,1.136);
\draw[gp path] (3.806,0.985)--(3.806,1.287);
\draw[gp path] (3.813,0.985)--(3.813,1.136);
\draw[gp path] (3.817,0.985)--(3.817,1.287);
\draw[gp path] (3.835,0.985)--(3.835,1.136);
\draw[gp path] (3.857,0.985)--(3.857,1.136);
\draw[gp path] (3.861,0.985)--(3.861,1.136);
\draw[gp path] (3.865,0.985)--(3.865,1.136);
\draw[gp path] (3.879,0.985)--(3.879,1.136);
\draw[gp path] (3.887,0.985)--(3.887,1.136);
\draw[gp path] (3.891,0.985)--(3.891,1.136);
\draw[gp path] (3.905,0.985)--(3.905,1.136);
\draw[gp path] (3.920,0.985)--(3.920,1.136);
\draw[gp path] (3.953,0.985)--(3.953,1.136);
\draw[gp path] (3.957,0.985)--(3.957,1.136);
\draw[gp path] (3.972,0.985)--(3.972,1.136);
\draw[gp path] (4.001,0.985)--(4.001,1.136);
\draw[gp path] (4.027,0.985)--(4.027,1.136);
\draw[gp path] (4.042,0.985)--(4.042,1.136);
\draw[gp path] (4.046,0.985)--(4.046,1.136);
\draw[gp path] (4.049,0.985)--(4.049,1.136);
\draw[gp path] (4.053,0.985)--(4.053,1.136);
\draw[gp path] (4.068,0.985)--(4.068,1.136);
\draw[gp path] (4.075,0.985)--(4.075,1.287);
\draw[gp path] (4.086,0.985)--(4.086,1.136);
\draw[gp path] (4.094,0.985)--(4.094,1.136);
\draw[gp path] (4.097,0.985)--(4.097,1.136);
\draw[gp path] (4.112,0.985)--(4.112,1.136);
\draw[gp path] (4.160,0.985)--(4.160,1.136);
\draw[gp path] (4.167,0.985)--(4.167,1.136);
\draw[gp path] (4.179,0.985)--(4.179,1.136);
\draw[gp path] (4.186,0.985)--(4.186,1.136);
\draw[gp path] (4.212,0.985)--(4.212,1.136);
\draw[gp path] (4.252,0.985)--(4.252,1.136);
\draw[gp path] (4.267,0.985)--(4.267,1.136);
\draw[gp path] (4.289,0.985)--(4.289,1.136);
\draw[gp path] (4.330,0.985)--(4.330,1.136);
\draw[gp path] (4.334,0.985)--(4.334,1.136);
\draw[gp path] (4.356,0.985)--(4.356,1.136);
\draw[gp path] (4.359,0.985)--(4.359,1.136);
\draw[gp path] (4.367,0.985)--(4.367,1.136);
\draw[gp path] (4.371,0.985)--(4.371,1.136);
\draw[gp path] (4.378,0.985)--(4.378,1.136);
\draw[gp path] (4.385,0.985)--(4.385,1.136);
\draw[gp path] (4.404,0.985)--(4.404,1.136);
\draw[gp path] (4.407,0.985)--(4.407,1.136);
\draw[gp path] (4.426,0.985)--(4.426,1.136);
\draw[gp path] (4.441,0.985)--(4.441,1.136);
\draw[gp path] (4.448,0.985)--(4.448,1.287);
\draw[gp path] (4.467,0.985)--(4.467,1.136);
\draw[gp path] (4.492,0.985)--(4.492,1.136);
\draw[gp path] (4.500,0.985)--(4.500,1.136);
\draw[gp path] (4.529,0.985)--(4.529,1.136);
\draw[gp path] (4.533,0.985)--(4.533,1.136);
\draw[gp path] (4.566,0.985)--(4.566,1.136);
\draw[gp path] (4.570,0.985)--(4.570,1.136);
\draw[gp path] (4.574,0.985)--(4.574,1.136);
\draw[gp path] (4.577,0.985)--(4.577,1.136);
\draw[gp path] (4.581,0.985)--(4.581,1.136);
\draw[gp path] (4.592,0.985)--(4.592,1.287);
\draw[gp path] (4.607,0.985)--(4.607,1.136);
\draw[gp path] (4.614,0.985)--(4.614,1.136);
\draw[gp path] (4.618,0.985)--(4.618,1.136);
\draw[gp path] (4.622,0.985)--(4.622,1.136);
\draw[gp path] (4.633,0.985)--(4.633,1.287);
\draw[gp path] (4.673,0.985)--(4.673,1.136);
\draw[gp path] (4.710,0.985)--(4.710,1.136);
\draw[gp path] (4.714,0.985)--(4.714,1.136);
\draw[gp path] (4.751,0.985)--(4.751,1.136);
\draw[gp path] (4.754,0.985)--(4.754,1.136);
\draw[gp path] (4.758,0.985)--(4.758,1.136);
\draw[gp path] (4.769,0.985)--(4.769,1.136);
\draw[gp path] (4.773,0.985)--(4.773,1.136);
\draw[gp path] (4.777,0.985)--(4.777,1.287);
\draw[gp path] (4.780,0.985)--(4.780,1.136);
\draw[gp path] (4.788,0.985)--(4.788,1.136);
\draw[gp path] (4.791,0.985)--(4.791,1.136);
\draw[gp path] (4.799,0.985)--(4.799,1.136);
\draw[gp path] (4.802,0.985)--(4.802,1.287);
\draw[gp path] (4.806,0.985)--(4.806,1.136);
\draw[gp path] (4.828,0.985)--(4.828,1.136);
\draw[gp path] (4.832,0.985)--(4.832,1.136);
\draw[gp path] (4.836,0.985)--(4.836,1.136);
\draw[gp path] (4.850,0.985)--(4.850,1.287);
\draw[gp path] (4.854,0.985)--(4.854,1.287);
\draw[gp path] (4.858,0.985)--(4.858,1.136);
\draw[gp path] (4.898,0.985)--(4.898,1.136);
\draw[gp path] (4.910,0.985)--(4.910,1.136);
\draw[gp path] (4.917,0.985)--(4.917,1.136);
\draw[gp path] (4.932,0.985)--(4.932,1.136);
\draw[gp path] (4.939,0.985)--(4.939,1.136);
\draw[gp path] (4.950,0.985)--(4.950,1.136);
\draw[gp path] (4.965,0.985)--(4.965,1.136);
\draw[gp path] (4.969,0.985)--(4.969,1.287);
\draw[gp path] (4.972,0.985)--(4.972,1.136);
\draw[gp path] (5.028,0.985)--(5.028,1.136);
\draw[gp path] (5.031,0.985)--(5.031,1.136);
\draw[gp path] (5.046,0.985)--(5.046,1.136);
\draw[gp path] (5.050,0.985)--(5.050,1.136);
\draw[gp path] (5.061,0.985)--(5.061,1.136);
\gpcolor{gp lt color border}
\node[gp node right] at (3.604,3.365) {$\rho^+$};
\gpcolor{gp lt color 1}
\gpsetlinetype{gp lt plot 1}
\draw[gp path] (3.788,3.365)--(4.704,3.365);
\draw[gp path] (1.380,2.043)--(1.417,2.043)--(1.455,2.043)--(1.492,2.043)--(1.529,2.043)%
  --(1.566,2.043)--(1.604,2.043)--(1.641,2.043)--(1.678,2.043)--(1.715,2.043)--(1.753,2.043)%
  --(1.790,2.043)--(1.827,2.043)--(1.864,2.043)--(1.902,2.043)--(1.939,2.043)--(1.976,2.043)%
  --(2.013,2.043)--(2.051,2.043)--(2.088,2.043)--(2.125,2.043)--(2.162,2.043)--(2.200,2.043)%
  --(2.237,2.043)--(2.274,2.043)--(2.311,2.043)--(2.349,2.043)--(2.386,2.043)--(2.423,2.043)%
  --(2.460,2.043)--(2.498,2.043)--(2.535,2.043)--(2.572,2.043)--(2.609,2.043)--(2.647,2.043)%
  --(2.684,2.043)--(2.721,2.043)--(2.758,2.043)--(2.796,2.043)--(2.833,2.043)--(2.870,2.043)%
  --(2.907,2.043)--(2.945,2.043)--(2.982,2.043)--(3.019,2.043)--(3.057,2.043)--(3.094,2.043)%
  --(3.131,2.043)--(3.168,2.043)--(3.206,2.043)--(3.243,2.043)--(3.280,2.043)--(3.317,2.043)%
  --(3.355,2.043)--(3.392,2.043)--(3.429,2.043)--(3.466,2.043)--(3.504,2.043)--(3.541,2.043)%
  --(3.578,2.043)--(3.615,2.043)--(3.653,2.043)--(3.690,2.043)--(3.727,2.043)--(3.764,2.043)%
  --(3.802,2.043)--(3.839,2.043)--(3.876,2.043)--(3.913,2.043)--(3.951,2.043)--(3.988,2.043)%
  --(4.025,2.043)--(4.062,2.043)--(4.100,2.043)--(4.137,2.043)--(4.174,2.043)--(4.211,2.043)%
  --(4.249,2.043)--(4.286,2.043)--(4.323,2.043)--(4.360,2.043)--(4.398,2.043)--(4.435,2.043)%
  --(4.472,2.043)--(4.509,2.043)--(4.547,2.043)--(4.584,2.043)--(4.621,2.043)--(4.658,2.043)%
  --(4.696,2.043)--(4.733,2.043)--(4.770,2.043)--(4.808,2.043)--(4.845,2.043)--(4.882,2.043)%
  --(4.919,2.043)--(4.957,2.043)--(4.994,2.043)--(5.031,2.043)--(5.068,2.043);
\gpcolor{gp lt color border}
\node[gp node right] at (3.604,3.057) {$\rho$};
\gpcolor{gp lt color 2}
\gpsetlinetype{gp lt plot 2}
\draw[gp path] (3.788,3.057)--(4.704,3.057);
\draw[gp path] (1.380,3.082)--(1.384,2.514)--(1.387,2.274)--(1.391,2.271)--(1.395,2.189)%
  --(1.398,2.207)--(1.402,2.200)--(1.406,2.144)--(1.410,2.164)--(1.413,2.188)--(1.417,2.235)%
  --(1.421,2.197)--(1.424,2.226)--(1.428,2.216)--(1.432,2.227)--(1.435,2.206)--(1.439,2.224)%
  --(1.443,2.254)--(1.446,2.269)--(1.450,2.301)--(1.454,2.343)--(1.458,2.283)--(1.461,2.352)%
  --(1.465,2.334)--(1.469,2.355)--(1.472,2.284)--(1.476,2.257)--(1.480,2.274)--(1.483,2.271)%
  --(1.487,2.342)--(1.491,2.324)--(1.494,2.345)--(1.498,2.278)--(1.502,2.362)--(1.506,2.319)%
  --(1.509,2.301)--(1.513,2.286)--(1.517,2.324)--(1.520,2.336)--(1.524,2.297)--(1.528,2.269)%
  --(1.531,2.303)--(1.535,2.294)--(1.539,2.291)--(1.542,2.272)--(1.546,2.289)--(1.550,2.337)%
  --(1.554,2.355)--(1.557,2.366)--(1.561,2.352)--(1.565,2.420)--(1.568,2.374)--(1.572,2.430)%
  --(1.576,2.310)--(1.579,2.392)--(1.583,2.399)--(1.587,2.434)--(1.590,2.470)--(1.594,2.487)%
  --(1.598,2.491)--(1.602,2.516)--(1.605,2.411)--(1.609,2.452)--(1.613,2.410)--(1.616,2.375)%
  --(1.620,2.437)--(1.624,2.404)--(1.627,2.368)--(1.631,2.395)--(1.635,2.340)--(1.638,2.334)%
  --(1.642,2.277)--(1.646,2.337)--(1.650,2.318)--(1.653,2.319)--(1.657,2.289)--(1.661,2.330)%
  --(1.664,2.304)--(1.668,2.333)--(1.672,2.309)--(1.675,2.291)--(1.679,2.339)--(1.683,2.318)%
  --(1.686,2.396)--(1.690,2.355)--(1.694,2.381)--(1.698,2.321)--(1.701,2.392)--(1.705,2.327)%
  --(1.709,2.390)--(1.712,2.307)--(1.716,2.351)--(1.720,2.334)--(1.723,2.322)--(1.727,2.265)%
  --(1.731,2.283)--(1.734,2.210)--(1.738,2.200)--(1.742,2.132)--(1.746,2.145)--(1.749,2.111)%
  --(1.753,2.114)--(1.757,2.108)--(1.760,2.129)--(1.764,2.096)--(1.768,2.127)--(1.771,2.111)%
  --(1.775,2.100)--(1.779,2.090)--(1.782,2.105)--(1.786,2.068)--(1.790,2.135)--(1.794,2.061)%
  --(1.797,2.067)--(1.801,2.052)--(1.805,1.972)--(1.808,1.935)--(1.812,1.926)--(1.816,1.911)%
  --(1.819,1.941)--(1.823,1.911)--(1.827,1.893)--(1.830,1.922)--(1.834,1.913)--(1.838,1.910)%
  --(1.842,1.961)--(1.845,1.925)--(1.849,1.949)--(1.853,1.907)--(1.856,1.957)--(1.860,1.955)%
  --(1.864,1.911)--(1.867,1.937)--(1.871,1.851)--(1.875,1.857)--(1.878,1.812)--(1.882,1.875)%
  --(1.886,1.867)--(1.889,1.878)--(1.893,1.880)--(1.897,1.896)--(1.901,1.908)--(1.904,1.931)%
  --(1.908,1.954)--(1.912,1.975)--(1.915,1.996)--(1.919,1.997)--(1.923,1.999)--(1.926,2.022)%
  --(1.930,1.978)--(1.934,2.050)--(1.937,1.972)--(1.941,1.940)--(1.945,1.914)--(1.949,1.934)%
  --(1.952,1.925)--(1.956,1.985)--(1.960,1.940)--(1.963,1.926)--(1.967,1.955)--(1.971,1.935)%
  --(1.974,1.991)--(1.978,1.985)--(1.982,2.002)--(1.985,2.044)--(1.989,1.984)--(1.993,2.008)%
  --(1.997,2.003)--(2.000,2.026)--(2.004,2.025)--(2.008,2.034)--(2.011,2.043)--(2.015,2.056)%
  --(2.019,2.112)--(2.022,2.091)--(2.026,2.158)--(2.030,2.046)--(2.033,2.077)--(2.037,2.068)%
  --(2.041,2.000)--(2.045,2.038)--(2.048,2.043)--(2.052,2.082)--(2.056,2.061)--(2.059,2.106)%
  --(2.063,2.067)--(2.067,2.052)--(2.070,2.103)--(2.074,2.093)--(2.078,2.133)--(2.081,2.145)%
  --(2.085,2.148)--(2.089,2.167)--(2.093,2.153)--(2.096,2.200)--(2.100,2.239)--(2.104,2.185)%
  --(2.107,2.192)--(2.111,2.207)--(2.115,2.271)--(2.118,2.238)--(2.122,2.265)--(2.126,2.294)%
  --(2.129,2.349)--(2.133,2.352)--(2.137,2.392)--(2.141,2.363)--(2.144,2.324)--(2.148,2.251)%
  --(2.152,2.265)--(2.155,2.283)--(2.159,2.334)--(2.163,2.301)--(2.166,2.387)--(2.170,2.366)%
  --(2.174,2.428)--(2.177,2.449)--(2.181,2.408)--(2.185,2.358)--(2.189,2.358)--(2.192,2.345)%
  --(2.196,2.337)--(2.200,2.316)--(2.203,2.357)--(2.207,2.312)--(2.211,2.298)--(2.214,2.310)%
  --(2.218,2.352)--(2.222,2.327)--(2.225,2.287)--(2.229,2.292)--(2.233,2.266)--(2.237,2.241)%
  --(2.240,2.194)--(2.244,2.230)--(2.248,2.173)--(2.251,2.291)--(2.255,2.200)--(2.259,2.212)%
  --(2.262,2.183)--(2.266,2.207)--(2.270,2.191)--(2.273,2.200)--(2.277,2.168)--(2.281,2.204)%
  --(2.285,2.223)--(2.288,2.177)--(2.292,2.218)--(2.296,2.135)--(2.299,2.192)--(2.303,2.105)%
  --(2.307,2.126)--(2.310,2.080)--(2.314,2.135)--(2.318,2.136)--(2.321,2.210)--(2.325,2.197)%
  --(2.329,2.213)--(2.333,2.177)--(2.336,2.232)--(2.340,2.215)--(2.344,2.250)--(2.347,2.229)%
  --(2.351,2.247)--(2.355,2.268)--(2.358,2.266)--(2.362,2.235)--(2.366,2.198)--(2.369,2.150)%
  --(2.373,2.216)--(2.377,2.127)--(2.381,2.135)--(2.384,2.153)--(2.388,2.188)--(2.392,2.136)%
  --(2.395,2.111)--(2.399,2.085)--(2.403,2.046)--(2.406,2.080)--(2.410,2.012)--(2.414,2.065)%
  --(2.417,2.064)--(2.421,2.097)--(2.425,2.126)--(2.429,2.111)--(2.432,2.155)--(2.436,2.207)%
  --(2.440,2.139)--(2.443,2.108)--(2.447,2.153)--(2.451,2.174)--(2.454,2.144)--(2.458,2.176)%
  --(2.462,2.176)--(2.465,2.235)--(2.469,2.224)--(2.473,2.244)--(2.477,2.306)--(2.480,2.200)%
  --(2.484,2.316)--(2.488,2.239)--(2.491,2.230)--(2.495,2.215)--(2.499,2.232)--(2.502,2.257)%
  --(2.506,2.294)--(2.510,2.268)--(2.513,2.238)--(2.517,2.242)--(2.521,2.215)--(2.525,2.188)%
  --(2.528,2.212)--(2.532,2.235)--(2.536,2.272)--(2.539,2.250)--(2.543,2.274)--(2.547,2.304)%
  --(2.550,2.260)--(2.554,2.303)--(2.558,2.207)--(2.561,2.186)--(2.565,2.213)--(2.569,2.236)%
  --(2.573,2.242)--(2.576,2.247)--(2.580,2.262)--(2.584,2.257)--(2.587,2.271)--(2.591,2.259)%
  --(2.595,2.219)--(2.598,2.281)--(2.602,2.339)--(2.606,2.333)--(2.609,2.306)--(2.613,2.419)%
  --(2.617,2.268)--(2.621,2.185)--(2.624,2.168)--(2.628,2.142)--(2.632,2.162)--(2.635,2.139)%
  --(2.639,2.162)--(2.643,2.162)--(2.646,2.145)--(2.650,2.218)--(2.654,2.221)--(2.657,2.223)%
  --(2.661,2.219)--(2.665,2.275)--(2.669,2.259)--(2.672,2.342)--(2.676,2.295)--(2.680,2.351)%
  --(2.683,2.318)--(2.687,2.303)--(2.691,2.304)--(2.694,2.245)--(2.698,2.239)--(2.702,2.263)%
  --(2.705,2.183)--(2.709,2.269)--(2.713,2.259)--(2.717,2.327)--(2.720,2.254)--(2.724,2.352)%
  --(2.728,2.304)--(2.731,2.369)--(2.735,2.322)--(2.739,2.389)--(2.742,2.322)--(2.746,2.336)%
  --(2.750,2.313)--(2.753,2.336)--(2.757,2.375)--(2.761,2.377)--(2.765,2.358)--(2.768,2.402)%
  --(2.772,2.358)--(2.776,2.303)--(2.779,2.371)--(2.783,2.417)--(2.787,2.439)--(2.790,2.352)%
  --(2.794,2.346)--(2.798,2.253)--(2.801,2.315)--(2.805,2.235)--(2.809,2.260)--(2.812,2.201)%
  --(2.816,2.159)--(2.820,2.145)--(2.824,2.117)--(2.827,2.067)--(2.831,2.085)--(2.835,2.067)%
  --(2.838,2.050)--(2.842,2.074)--(2.846,2.117)--(2.849,2.058)--(2.853,2.096)--(2.857,2.043)%
  --(2.860,2.074)--(2.864,2.079)--(2.868,2.070)--(2.872,2.017)--(2.875,2.012)--(2.879,1.993)%
  --(2.883,1.973)--(2.886,1.925)--(2.890,1.931)--(2.894,1.899)--(2.897,1.893)--(2.901,1.836)%
  --(2.905,1.875)--(2.908,1.821)--(2.912,1.837)--(2.916,1.846)--(2.920,1.858)--(2.923,1.893)%
  --(2.927,1.893)--(2.931,1.919)--(2.934,1.916)--(2.938,1.858)--(2.942,1.878)--(2.945,1.861)%
  --(2.949,1.873)--(2.953,1.913)--(2.956,1.858)--(2.960,1.883)--(2.964,1.901)--(2.968,1.931)%
  --(2.971,1.886)--(2.975,1.884)--(2.979,1.922)--(2.982,1.999)--(2.986,2.019)--(2.990,2.068)%
  --(2.993,2.035)--(2.997,2.088)--(3.001,2.118)--(3.004,2.141)--(3.008,2.123)--(3.012,2.168)%
  --(3.016,2.085)--(3.019,2.115)--(3.023,2.100)--(3.027,2.167)--(3.030,2.185)--(3.034,2.174)%
  --(3.038,2.174)--(3.041,2.189)--(3.045,2.188)--(3.049,2.186)--(3.052,2.216)--(3.056,2.165)%
  --(3.060,2.174)--(3.064,2.167)--(3.067,2.123)--(3.071,2.164)--(3.075,2.126)--(3.078,2.170)%
  --(3.082,2.138)--(3.086,2.141)--(3.089,2.059)--(3.093,2.049)--(3.097,2.062)--(3.100,2.087)%
  --(3.104,2.070)--(3.108,2.085)--(3.112,2.082)--(3.115,2.097)--(3.119,2.083)--(3.123,2.120)%
  --(3.126,2.083)--(3.130,2.150)--(3.134,2.108)--(3.137,2.142)--(3.141,2.161)--(3.145,2.221)%
  --(3.148,2.141)--(3.152,2.085)--(3.156,2.068)--(3.160,2.097)--(3.163,2.090)--(3.167,2.133)%
  --(3.171,2.150)--(3.174,2.123)--(3.178,2.109)--(3.182,2.068)--(3.185,2.035)--(3.189,2.031)%
  --(3.193,1.999)--(3.196,1.973)--(3.200,1.964)--(3.204,2.003)--(3.208,1.990)--(3.211,2.043)%
  --(3.215,2.059)--(3.219,2.050)--(3.222,2.109)--(3.226,2.162)--(3.230,2.147)--(3.233,2.161)%
  --(3.237,2.161)--(3.241,2.209)--(3.244,2.150)--(3.248,2.156)--(3.252,2.145)--(3.256,2.123)%
  --(3.259,2.077)--(3.263,2.014)--(3.267,1.967)--(3.270,1.970)--(3.274,1.935)--(3.278,1.893)%
  --(3.281,1.934)--(3.285,1.904)--(3.289,1.851)--(3.292,1.812)--(3.296,1.833)--(3.300,1.816)%
  --(3.304,1.849)--(3.307,1.864)--(3.311,1.895)--(3.315,1.914)--(3.318,1.892)--(3.322,1.987)%
  --(3.326,1.961)--(3.329,1.993)--(3.333,1.978)--(3.337,2.047)--(3.340,1.975)--(3.344,2.061)%
  --(3.348,2.035)--(3.352,2.100)--(3.355,2.106)--(3.359,2.153)--(3.363,2.136)--(3.366,2.139)%
  --(3.370,2.186)--(3.374,2.115)--(3.377,2.174)--(3.381,2.090)--(3.385,2.077)--(3.388,2.102)%
  --(3.392,2.067)--(3.396,2.112)--(3.400,2.097)--(3.403,2.097)--(3.407,2.112)--(3.411,2.156)%
  --(3.414,2.138)--(3.418,2.176)--(3.422,2.174)--(3.425,2.164)--(3.429,2.133)--(3.433,2.139)%
  --(3.436,2.150)--(3.440,2.167)--(3.444,2.156)--(3.448,2.170)--(3.451,2.148)--(3.455,2.192)%
  --(3.459,2.183)--(3.462,2.162)--(3.466,2.159)--(3.470,2.179)--(3.473,2.191)--(3.477,2.230)%
  --(3.481,2.201)--(3.484,2.145)--(3.488,2.269)--(3.492,2.168)--(3.496,2.186)--(3.499,2.165)%
  --(3.503,2.176)--(3.507,2.147)--(3.510,2.170)--(3.514,2.156)--(3.518,2.167)--(3.521,2.106)%
  --(3.525,2.161)--(3.529,2.127)--(3.532,2.230)--(3.536,2.168)--(3.540,2.204)--(3.544,2.191)%
  --(3.547,2.218)--(3.551,2.233)--(3.555,2.198)--(3.558,2.156)--(3.562,2.148)--(3.566,2.144)%
  --(3.569,2.076)--(3.573,2.091)--(3.577,2.096)--(3.580,2.120)--(3.584,2.102)--(3.588,2.121)%
  --(3.592,2.188)--(3.595,2.174)--(3.599,2.138)--(3.603,2.133)--(3.606,2.150)--(3.610,2.158)%
  --(3.614,2.168)--(3.617,2.114)--(3.621,2.171)--(3.625,2.130)--(3.628,2.182)--(3.632,2.090)%
  --(3.636,2.164)--(3.640,2.070)--(3.643,2.088)--(3.647,2.043)--(3.651,2.096)--(3.654,2.043)%
  --(3.658,2.061)--(3.662,2.019)--(3.665,2.009)--(3.669,2.016)--(3.673,2.022)--(3.676,2.020)%
  --(3.680,2.026)--(3.684,2.032)--(3.688,2.052)--(3.691,2.071)--(3.695,2.059)--(3.699,2.108)%
  --(3.702,2.153)--(3.706,2.150)--(3.710,2.114)--(3.713,2.150)--(3.717,2.102)--(3.721,2.182)%
  --(3.724,2.194)--(3.728,2.180)--(3.732,2.162)--(3.735,2.265)--(3.739,2.191)--(3.743,2.176)%
  --(3.747,2.219)--(3.750,2.179)--(3.754,2.257)--(3.758,2.200)--(3.761,2.301)--(3.765,2.212)%
  --(3.769,2.303)--(3.772,2.327)--(3.776,2.410)--(3.780,2.365)--(3.783,2.390)--(3.787,2.358)%
  --(3.791,2.349)--(3.795,2.325)--(3.798,2.426)--(3.802,2.369)--(3.806,2.313)--(3.809,2.313)%
  --(3.813,2.304)--(3.817,2.298)--(3.820,2.340)--(3.824,2.253)--(3.828,2.313)--(3.831,2.321)%
  --(3.835,2.315)--(3.839,2.333)--(3.843,2.307)--(3.846,2.277)--(3.850,2.292)--(3.854,2.294)%
  --(3.857,2.268)--(3.861,2.242)--(3.865,2.218)--(3.868,2.224)--(3.872,2.232)--(3.876,2.236)%
  --(3.879,2.189)--(3.883,2.133)--(3.887,2.144)--(3.891,2.087)--(3.894,2.120)--(3.898,2.082)%
  --(3.902,2.100)--(3.905,2.118)--(3.909,2.083)--(3.913,2.111)--(3.916,2.056)--(3.920,2.124)%
  --(3.924,2.132)--(3.927,2.138)--(3.931,2.130)--(3.935,2.118)--(3.939,2.106)--(3.942,2.138)%
  --(3.946,2.135)--(3.950,2.197)--(3.953,2.171)--(3.957,2.136)--(3.961,2.115)--(3.964,2.147)%
  --(3.968,2.121)--(3.972,2.088)--(3.975,2.105)--(3.979,2.097)--(3.983,2.150)--(3.987,2.067)%
  --(3.990,2.156)--(3.994,2.105)--(3.998,2.195)--(4.001,2.111)--(4.005,2.185)--(4.009,2.155)%
  --(4.012,2.232)--(4.016,2.207)--(4.020,2.123)--(4.023,2.109)--(4.027,2.032)--(4.031,2.056)%
  --(4.035,2.026)--(4.038,2.044)--(4.042,2.038)--(4.046,2.017)--(4.049,2.044)--(4.053,2.016)%
  --(4.057,2.055)--(4.060,1.988)--(4.064,2.041)--(4.068,2.019)--(4.071,2.082)--(4.075,2.082)%
  --(4.079,2.133)--(4.083,2.123)--(4.086,2.198)--(4.090,2.148)--(4.094,2.133)--(4.097,2.114)%
  --(4.101,2.189)--(4.105,2.168)--(4.108,2.186)--(4.112,2.221)--(4.116,2.180)--(4.119,2.229)%
  --(4.123,2.129)--(4.127,2.206)--(4.131,2.130)--(4.134,2.209)--(4.138,2.074)--(4.142,2.052)%
  --(4.145,2.108)--(4.149,2.091)--(4.153,2.068)--(4.156,2.094)--(4.160,2.102)--(4.164,2.074)%
  --(4.167,2.044)--(4.171,2.087)--(4.175,2.062)--(4.179,2.049)--(4.182,2.099)--(4.186,2.074)%
  --(4.190,2.082)--(4.193,2.087)--(4.197,2.073)--(4.201,2.091)--(4.204,2.126)--(4.208,2.114)%
  --(4.212,2.139)--(4.215,2.133)--(4.219,2.183)--(4.223,2.135)--(4.227,2.165)--(4.230,2.094)%
  --(4.234,2.114)--(4.238,2.079)--(4.241,2.130)--(4.245,2.067)--(4.249,2.103)--(4.252,2.077)%
  --(4.256,2.061)--(4.260,2.068)--(4.263,2.053)--(4.267,2.014)--(4.271,1.997)--(4.275,2.003)%
  --(4.278,2.005)--(4.282,2.017)--(4.286,2.012)--(4.289,2.085)--(4.293,2.043)--(4.297,2.067)%
  --(4.300,2.005)--(4.304,1.996)--(4.308,2.017)--(4.311,2.005)--(4.315,2.022)--(4.319,2.011)%
  --(4.323,1.970)--(4.326,1.987)--(4.330,1.957)--(4.334,1.931)--(4.337,1.911)--(4.341,1.931)%
  --(4.345,1.967)--(4.348,1.975)--(4.352,2.009)--(4.356,1.991)--(4.359,1.981)--(4.363,2.005)%
  --(4.367,2.023)--(4.371,2.046)--(4.374,1.999)--(4.378,1.990)--(4.382,2.006)--(4.385,2.031)%
  --(4.389,2.064)--(4.393,2.103)--(4.396,2.162)--(4.400,2.164)--(4.404,2.200)--(4.407,2.191)%
  --(4.411,2.145)--(4.415,2.210)--(4.419,2.215)--(4.422,2.260)--(4.426,2.216)--(4.430,2.226)%
  --(4.433,2.238)--(4.437,2.185)--(4.441,2.180)--(4.444,2.182)--(4.448,2.171)--(4.452,2.141)%
  --(4.455,2.159)--(4.459,2.085)--(4.463,2.097)--(4.467,2.062)--(4.470,2.061)--(4.474,2.034)%
  --(4.478,2.047)--(4.481,2.058)--(4.485,2.037)--(4.489,2.055)--(4.492,2.046)--(4.496,2.070)%
  --(4.500,2.046)--(4.503,2.094)--(4.507,2.085)--(4.511,2.144)--(4.515,2.077)--(4.518,2.094)%
  --(4.522,2.120)--(4.526,2.115)--(4.529,2.135)--(4.533,2.097)--(4.537,2.129)--(4.540,2.091)%
  --(4.544,2.121)--(4.548,2.177)--(4.551,2.127)--(4.555,2.147)--(4.559,2.068)--(4.563,2.070)%
  --(4.566,2.032)--(4.570,2.068)--(4.574,2.041)--(4.577,2.062)--(4.581,2.040)--(4.585,2.064)%
  --(4.588,2.041)--(4.592,2.029)--(4.596,2.064)--(4.599,2.073)--(4.603,2.068)--(4.607,2.105)%
  --(4.611,2.102)--(4.614,2.132)--(4.618,2.094)--(4.622,2.090)--(4.625,2.112)--(4.629,2.139)%
  --(4.633,2.058)--(4.636,2.147)--(4.640,2.136)--(4.644,2.153)--(4.647,2.127)--(4.651,2.103)%
  --(4.655,2.111)--(4.658,2.046)--(4.662,2.016)--(4.666,2.099)--(4.670,2.067)--(4.673,2.094)%
  --(4.677,2.094)--(4.681,2.067)--(4.684,2.091)--(4.688,2.077)--(4.692,2.162)--(4.695,2.156)%
  --(4.699,2.223)--(4.703,2.176)--(4.706,2.250)--(4.710,2.170)--(4.714,2.176)--(4.718,2.150)%
  --(4.721,2.148)--(4.725,2.133)--(4.729,2.167)--(4.732,2.106)--(4.736,2.121)--(4.740,2.118)%
  --(4.743,2.080)--(4.747,2.083)--(4.751,2.068)--(4.754,2.067)--(4.758,2.061)--(4.762,2.080)%
  --(4.766,2.062)--(4.769,2.070)--(4.773,2.082)--(4.777,2.044)--(4.780,2.032)--(4.784,1.990)%
  --(4.788,2.019)--(4.791,1.979)--(4.795,2.014)--(4.799,2.012)--(4.802,1.993)--(4.806,1.991)%
  --(4.810,1.997)--(4.814,1.978)--(4.817,2.002)--(4.821,2.002)--(4.825,2.005)--(4.828,2.014)%
  --(4.832,1.978)--(4.836,2.012)--(4.839,1.951)--(4.843,1.934)--(4.847,1.948)--(4.850,1.960)%
  --(4.854,1.899)--(4.858,1.886)--(4.862,1.843)--(4.865,1.883)--(4.869,1.836)--(4.873,1.833)%
  --(4.876,1.831)--(4.880,1.833)--(4.884,1.786)--(4.887,1.795)--(4.891,1.802)--(4.895,1.846)%
  --(4.898,1.830)--(4.902,1.839)--(4.906,1.872)--(4.910,1.872)--(4.913,1.861)--(4.917,1.824)%
  --(4.921,1.893)--(4.924,1.839)--(4.928,1.880)--(4.932,1.818)--(4.935,1.855)--(4.939,1.796)%
  --(4.943,1.816)--(4.946,1.821)--(4.950,1.883)--(4.954,1.843)--(4.958,1.880)--(4.961,1.861)%
  --(4.965,1.899)--(4.969,1.857)--(4.972,1.880)--(4.976,1.857)--(4.980,1.830)--(4.983,1.902)%
  --(4.987,1.896)--(4.991,1.840)--(4.994,1.855)--(4.998,1.869)--(5.002,1.860)--(5.006,1.892)%
  --(5.009,1.896)--(5.013,1.919)--(5.017,1.937)--(5.020,1.914)--(5.024,1.941)--(5.028,1.979)%
  --(5.031,1.911)--(5.035,1.964)--(5.039,1.952)--(5.042,2.000)--(5.046,1.994)--(5.050,2.062)%
  --(5.054,2.025)--(5.057,2.022)--(5.061,1.985)--(5.065,2.052)--(5.068,1.985);
\gpcolor{gp lt color border}
\gpsetlinetype{gp lt border}
\draw[gp path] (1.380,4.007)--(1.380,0.985)--(5.072,0.985)--(5.072,4.007)--cycle;
\gpdefrectangularnode{gp plot 1}{\pgfpoint{1.380cm}{0.985cm}}{\pgfpoint{5.072cm}{4.007cm}}
\end{tikzpicture}

%% file: figures/ml009.tex
\begin{tikzpicture}[gnuplot]
\gpmonochromelines
\gpcolor{gp lt color border}
\gpsetlinetype{gp lt border}
\gpsetlinewidth{1.00}
\draw[gp path] (1.380,0.985)--(1.560,0.985);
\draw[gp path] (5.072,0.985)--(4.892,0.985);
\node[gp node right] at (1.196,0.985) { 0};
\draw[gp path] (1.380,1.741)--(1.560,1.741);
\draw[gp path] (5.072,1.741)--(4.892,1.741);
\node[gp node right] at (1.196,1.741) { 0.05};
\draw[gp path] (1.380,2.496)--(1.560,2.496);
\draw[gp path] (5.072,2.496)--(4.892,2.496);
\node[gp node right] at (1.196,2.496) { 0.1};
\draw[gp path] (1.380,3.252)--(1.560,3.252);
\draw[gp path] (5.072,3.252)--(4.892,3.252);
\node[gp node right] at (1.196,3.252) { 0.15};
\draw[gp path] (1.380,4.007)--(1.560,4.007);
\draw[gp path] (5.072,4.007)--(4.892,4.007);
\node[gp node right] at (1.196,4.007) { 0.2};
\draw[gp path] (1.380,0.985)--(1.380,1.165);
\draw[gp path] (1.380,4.007)--(1.380,3.827);
\node[gp node center] at (1.380,0.677) { 0};
\draw[gp path] (2.118,0.985)--(2.118,1.165);
\draw[gp path] (2.118,4.007)--(2.118,3.827);
\node[gp node center] at (2.118,0.677) { 200};
\draw[gp path] (2.857,0.985)--(2.857,1.165);
\draw[gp path] (2.857,4.007)--(2.857,3.827);
\node[gp node center] at (2.857,0.677) { 400};
\draw[gp path] (3.595,0.985)--(3.595,1.165);
\draw[gp path] (3.595,4.007)--(3.595,3.827);
\node[gp node center] at (3.595,0.677) { 600};
\draw[gp path] (4.334,0.985)--(4.334,1.165);
\draw[gp path] (4.334,4.007)--(4.334,3.827);
\node[gp node center] at (4.334,0.677) { 800};
\draw[gp path] (5.072,0.985)--(5.072,1.165);
\draw[gp path] (5.072,4.007)--(5.072,3.827);
\node[gp node center] at (5.072,0.677) { 1000};
\draw[gp path] (1.380,4.007)--(1.380,0.985)--(5.072,0.985)--(5.072,4.007)--cycle;
\node[gp node center] at (3.226,0.215) {step};
\node[gp node right] at (3.604,3.673) {r (\%)};
\gpcolor{gp lt color 0}
\gpsetlinetype{gp lt plot 0}
\draw[gp path] (3.788,3.673)--(4.704,3.673);
\draw[gp path] (1.384,0.985)--(1.384,1.287);
\draw[gp path] (1.387,0.985)--(1.387,1.136);
\draw[gp path] (1.391,0.985)--(1.391,1.287);
\draw[gp path] (1.398,0.985)--(1.398,1.136);
\draw[gp path] (1.402,0.985)--(1.402,1.287);
\draw[gp path] (1.406,0.985)--(1.406,1.136);
\draw[gp path] (1.410,0.985)--(1.410,1.438);
\draw[gp path] (1.413,0.985)--(1.413,1.589);
\draw[gp path] (1.417,0.985)--(1.417,1.136);
\draw[gp path] (1.421,0.985)--(1.421,1.438);
\draw[gp path] (1.428,0.985)--(1.428,1.287);
\draw[gp path] (1.432,0.985)--(1.432,1.136);
\draw[gp path] (1.435,0.985)--(1.435,1.287);
\draw[gp path] (1.439,0.985)--(1.439,1.136);
\draw[gp path] (1.443,0.985)--(1.443,1.136);
\draw[gp path] (1.450,0.985)--(1.450,1.287);
\draw[gp path] (1.454,0.985)--(1.454,1.136);
\draw[gp path] (1.458,0.985)--(1.458,1.136);
\draw[gp path] (1.461,0.985)--(1.461,1.287);
\draw[gp path] (1.465,0.985)--(1.465,1.438);
\draw[gp path] (1.469,0.985)--(1.469,1.287);
\draw[gp path] (1.472,0.985)--(1.472,1.136);
\draw[gp path] (1.483,0.985)--(1.483,1.287);
\draw[gp path] (1.487,0.985)--(1.487,1.287);
\draw[gp path] (1.491,0.985)--(1.491,1.136);
\draw[gp path] (1.494,0.985)--(1.494,1.287);
\draw[gp path] (1.498,0.985)--(1.498,1.287);
\draw[gp path] (1.502,0.985)--(1.502,1.136);
\draw[gp path] (1.506,0.985)--(1.506,1.136);
\draw[gp path] (1.509,0.985)--(1.509,1.136);
\draw[gp path] (1.517,0.985)--(1.517,1.136);
\draw[gp path] (1.520,0.985)--(1.520,1.287);
\draw[gp path] (1.531,0.985)--(1.531,1.136);
\draw[gp path] (1.535,0.985)--(1.535,1.287);
\draw[gp path] (1.539,0.985)--(1.539,1.136);
\draw[gp path] (1.546,0.985)--(1.546,1.136);
\draw[gp path] (1.554,0.985)--(1.554,1.287);
\draw[gp path] (1.557,0.985)--(1.557,1.136);
\draw[gp path] (1.565,0.985)--(1.565,1.287);
\draw[gp path] (1.568,0.985)--(1.568,1.136);
\draw[gp path] (1.572,0.985)--(1.572,1.136);
\draw[gp path] (1.576,0.985)--(1.576,1.438);
\draw[gp path] (1.583,0.985)--(1.583,1.136);
\draw[gp path] (1.587,0.985)--(1.587,1.287);
\draw[gp path] (1.594,0.985)--(1.594,1.589);
\draw[gp path] (1.598,0.985)--(1.598,1.136);
\draw[gp path] (1.602,0.985)--(1.602,1.438);
\draw[gp path] (1.605,0.985)--(1.605,1.589);
\draw[gp path] (1.609,0.985)--(1.609,1.136);
\draw[gp path] (1.613,0.985)--(1.613,1.136);
\draw[gp path] (1.616,0.985)--(1.616,1.136);
\draw[gp path] (1.620,0.985)--(1.620,1.892);
\draw[gp path] (1.624,0.985)--(1.624,1.438);
\draw[gp path] (1.635,0.985)--(1.635,1.136);
\draw[gp path] (1.638,0.985)--(1.638,1.287);
\draw[gp path] (1.642,0.985)--(1.642,1.136);
\draw[gp path] (1.653,0.985)--(1.653,1.136);
\draw[gp path] (1.661,0.985)--(1.661,1.438);
\draw[gp path] (1.668,0.985)--(1.668,1.136);
\draw[gp path] (1.672,0.985)--(1.672,1.287);
\draw[gp path] (1.675,0.985)--(1.675,1.438);
\draw[gp path] (1.679,0.985)--(1.679,1.287);
\draw[gp path] (1.686,0.985)--(1.686,1.136);
\draw[gp path] (1.709,0.985)--(1.709,1.136);
\draw[gp path] (1.712,0.985)--(1.712,1.136);
\draw[gp path] (1.716,0.985)--(1.716,1.287);
\draw[gp path] (1.731,0.985)--(1.731,1.136);
\draw[gp path] (1.734,0.985)--(1.734,1.287);
\draw[gp path] (1.738,0.985)--(1.738,1.287);
\draw[gp path] (1.746,0.985)--(1.746,1.136);
\draw[gp path] (1.749,0.985)--(1.749,1.287);
\draw[gp path] (1.753,0.985)--(1.753,1.287);
\draw[gp path] (1.757,0.985)--(1.757,1.741);
\draw[gp path] (1.760,0.985)--(1.760,1.136);
\draw[gp path] (1.768,0.985)--(1.768,1.438);
\draw[gp path] (1.775,0.985)--(1.775,1.287);
\draw[gp path] (1.779,0.985)--(1.779,1.136);
\draw[gp path] (1.782,0.985)--(1.782,1.438);
\draw[gp path] (1.790,0.985)--(1.790,1.287);
\draw[gp path] (1.797,0.985)--(1.797,1.136);
\draw[gp path] (1.801,0.985)--(1.801,1.136);
\draw[gp path] (1.805,0.985)--(1.805,1.136);
\draw[gp path] (1.808,0.985)--(1.808,1.438);
\draw[gp path] (1.812,0.985)--(1.812,1.287);
\draw[gp path] (1.819,0.985)--(1.819,1.136);
\draw[gp path] (1.823,0.985)--(1.823,1.136);
\draw[gp path] (1.827,0.985)--(1.827,1.438);
\draw[gp path] (1.830,0.985)--(1.830,1.136);
\draw[gp path] (1.838,0.985)--(1.838,1.589);
\draw[gp path] (1.849,0.985)--(1.849,1.287);
\draw[gp path] (1.856,0.985)--(1.856,1.136);
\draw[gp path] (1.860,0.985)--(1.860,1.287);
\draw[gp path] (1.864,0.985)--(1.864,1.741);
\draw[gp path] (1.867,0.985)--(1.867,1.136);
\draw[gp path] (1.878,0.985)--(1.878,1.438);
\draw[gp path] (1.889,0.985)--(1.889,1.438);
\draw[gp path] (1.893,0.985)--(1.893,1.287);
\draw[gp path] (1.897,0.985)--(1.897,1.287);
\draw[gp path] (1.901,0.985)--(1.901,1.136);
\draw[gp path] (1.908,0.985)--(1.908,1.287);
\draw[gp path] (1.912,0.985)--(1.912,1.136);
\draw[gp path] (1.915,0.985)--(1.915,1.136);
\draw[gp path] (1.923,0.985)--(1.923,1.136);
\draw[gp path] (1.926,0.985)--(1.926,1.136);
\draw[gp path] (1.930,0.985)--(1.930,1.287);
\draw[gp path] (1.934,0.985)--(1.934,1.136);
\draw[gp path] (1.941,0.985)--(1.941,1.136);
\draw[gp path] (1.949,0.985)--(1.949,1.136);
\draw[gp path] (1.952,0.985)--(1.952,1.136);
\draw[gp path] (1.956,0.985)--(1.956,1.136);
\draw[gp path] (1.963,0.985)--(1.963,1.287);
\draw[gp path] (1.971,0.985)--(1.971,1.136);
\draw[gp path] (1.974,0.985)--(1.974,1.136);
\draw[gp path] (1.978,0.985)--(1.978,1.136);
\draw[gp path] (1.982,0.985)--(1.982,1.589);
\draw[gp path] (1.985,0.985)--(1.985,1.287);
\draw[gp path] (1.989,0.985)--(1.989,1.438);
\draw[gp path] (1.997,0.985)--(1.997,1.438);
\draw[gp path] (2.000,0.985)--(2.000,1.287);
\draw[gp path] (2.004,0.985)--(2.004,1.136);
\draw[gp path] (2.008,0.985)--(2.008,1.136);
\draw[gp path] (2.011,0.985)--(2.011,1.136);
\draw[gp path] (2.015,0.985)--(2.015,1.136);
\draw[gp path] (2.030,0.985)--(2.030,1.287);
\draw[gp path] (2.041,0.985)--(2.041,1.136);
\draw[gp path] (2.048,0.985)--(2.048,1.136);
\draw[gp path] (2.052,0.985)--(2.052,1.287);
\draw[gp path] (2.059,0.985)--(2.059,1.136);
\draw[gp path] (2.063,0.985)--(2.063,1.287);
\draw[gp path] (2.074,0.985)--(2.074,1.136);
\draw[gp path] (2.081,0.985)--(2.081,1.136);
\draw[gp path] (2.089,0.985)--(2.089,1.136);
\draw[gp path] (2.096,0.985)--(2.096,1.136);
\draw[gp path] (2.100,0.985)--(2.100,1.136);
\draw[gp path] (2.104,0.985)--(2.104,1.136);
\draw[gp path] (2.107,0.985)--(2.107,1.438);
\draw[gp path] (2.111,0.985)--(2.111,1.287);
\draw[gp path] (2.115,0.985)--(2.115,1.287);
\draw[gp path] (2.118,0.985)--(2.118,1.136);
\draw[gp path] (2.122,0.985)--(2.122,1.136);
\draw[gp path] (2.126,0.985)--(2.126,1.136);
\draw[gp path] (2.129,0.985)--(2.129,1.287);
\draw[gp path] (2.133,0.985)--(2.133,1.136);
\draw[gp path] (2.141,0.985)--(2.141,1.136);
\draw[gp path] (2.148,0.985)--(2.148,1.136);
\draw[gp path] (2.152,0.985)--(2.152,1.287);
\draw[gp path] (2.155,0.985)--(2.155,1.136);
\draw[gp path] (2.163,0.985)--(2.163,1.136);
\draw[gp path] (2.174,0.985)--(2.174,1.136);
\draw[gp path] (2.177,0.985)--(2.177,1.287);
\draw[gp path] (2.185,0.985)--(2.185,1.136);
\draw[gp path] (2.189,0.985)--(2.189,1.287);
\draw[gp path] (2.196,0.985)--(2.196,1.136);
\draw[gp path] (2.203,0.985)--(2.203,1.287);
\draw[gp path] (2.207,0.985)--(2.207,1.136);
\draw[gp path] (2.211,0.985)--(2.211,1.287);
\draw[gp path] (2.214,0.985)--(2.214,1.136);
\draw[gp path] (2.222,0.985)--(2.222,1.136);
\draw[gp path] (2.225,0.985)--(2.225,1.136);
\draw[gp path] (2.229,0.985)--(2.229,1.287);
\draw[gp path] (2.237,0.985)--(2.237,1.136);
\draw[gp path] (2.244,0.985)--(2.244,1.136);
\draw[gp path] (2.251,0.985)--(2.251,1.287);
\draw[gp path] (2.255,0.985)--(2.255,1.589);
\draw[gp path] (2.270,0.985)--(2.270,1.438);
\draw[gp path] (2.273,0.985)--(2.273,1.136);
\draw[gp path] (2.277,0.985)--(2.277,1.438);
\draw[gp path] (2.281,0.985)--(2.281,1.136);
\draw[gp path] (2.288,0.985)--(2.288,1.136);
\draw[gp path] (2.292,0.985)--(2.292,1.287);
\draw[gp path] (2.296,0.985)--(2.296,1.136);
\draw[gp path] (2.299,0.985)--(2.299,1.136);
\draw[gp path] (2.307,0.985)--(2.307,1.438);
\draw[gp path] (2.310,0.985)--(2.310,1.438);
\draw[gp path] (2.314,0.985)--(2.314,1.438);
\draw[gp path] (2.325,0.985)--(2.325,1.136);
\draw[gp path] (2.329,0.985)--(2.329,1.136);
\draw[gp path] (2.333,0.985)--(2.333,1.136);
\draw[gp path] (2.336,0.985)--(2.336,1.136);
\draw[gp path] (2.344,0.985)--(2.344,1.287);
\draw[gp path] (2.347,0.985)--(2.347,1.136);
\draw[gp path] (2.355,0.985)--(2.355,1.287);
\draw[gp path] (2.358,0.985)--(2.358,1.136);
\draw[gp path] (2.362,0.985)--(2.362,1.136);
\draw[gp path] (2.369,0.985)--(2.369,1.136);
\draw[gp path] (2.388,0.985)--(2.388,1.136);
\draw[gp path] (2.395,0.985)--(2.395,1.136);
\draw[gp path] (2.399,0.985)--(2.399,1.136);
\draw[gp path] (2.406,0.985)--(2.406,1.438);
\draw[gp path] (2.410,0.985)--(2.410,1.136);
\draw[gp path] (2.417,0.985)--(2.417,1.287);
\draw[gp path] (2.421,0.985)--(2.421,1.287);
\draw[gp path] (2.425,0.985)--(2.425,1.438);
\draw[gp path] (2.429,0.985)--(2.429,1.136);
\draw[gp path] (2.440,0.985)--(2.440,1.136);
\draw[gp path] (2.451,0.985)--(2.451,1.136);
\draw[gp path] (2.454,0.985)--(2.454,1.136);
\draw[gp path] (2.458,0.985)--(2.458,1.438);
\draw[gp path] (2.465,0.985)--(2.465,1.136);
\draw[gp path] (2.469,0.985)--(2.469,1.136);
\draw[gp path] (2.473,0.985)--(2.473,1.136);
\draw[gp path] (2.477,0.985)--(2.477,1.136);
\draw[gp path] (2.480,0.985)--(2.480,1.136);
\draw[gp path] (2.484,0.985)--(2.484,1.136);
\draw[gp path] (2.495,0.985)--(2.495,1.287);
\draw[gp path] (2.499,0.985)--(2.499,1.136);
\draw[gp path] (2.502,0.985)--(2.502,1.136);
\draw[gp path] (2.513,0.985)--(2.513,1.136);
\draw[gp path] (2.517,0.985)--(2.517,1.287);
\draw[gp path] (2.525,0.985)--(2.525,1.136);
\draw[gp path] (2.528,0.985)--(2.528,1.287);
\draw[gp path] (2.532,0.985)--(2.532,1.136);
\draw[gp path] (2.536,0.985)--(2.536,1.136);
\draw[gp path] (2.539,0.985)--(2.539,1.136);
\draw[gp path] (2.543,0.985)--(2.543,1.136);
\draw[gp path] (2.547,0.985)--(2.547,1.136);
\draw[gp path] (2.550,0.985)--(2.550,1.287);
\draw[gp path] (2.554,0.985)--(2.554,1.136);
\draw[gp path] (2.565,0.985)--(2.565,1.136);
\draw[gp path] (2.569,0.985)--(2.569,1.136);
\draw[gp path] (2.573,0.985)--(2.573,1.136);
\draw[gp path] (2.576,0.985)--(2.576,1.136);
\draw[gp path] (2.580,0.985)--(2.580,1.136);
\draw[gp path] (2.584,0.985)--(2.584,1.287);
\draw[gp path] (2.595,0.985)--(2.595,1.287);
\draw[gp path] (2.598,0.985)--(2.598,1.287);
\draw[gp path] (2.602,0.985)--(2.602,1.438);
\draw[gp path] (2.606,0.985)--(2.606,1.438);
\draw[gp path] (2.609,0.985)--(2.609,1.136);
\draw[gp path] (2.613,0.985)--(2.613,1.589);
\draw[gp path] (2.617,0.985)--(2.617,1.136);
\draw[gp path] (2.621,0.985)--(2.621,1.136);
\draw[gp path] (2.624,0.985)--(2.624,1.136);
\draw[gp path] (2.632,0.985)--(2.632,1.589);
\draw[gp path] (2.635,0.985)--(2.635,1.287);
\draw[gp path] (2.639,0.985)--(2.639,1.287);
\draw[gp path] (2.654,0.985)--(2.654,1.136);
\draw[gp path] (2.657,0.985)--(2.657,1.136);
\draw[gp path] (2.661,0.985)--(2.661,1.438);
\draw[gp path] (2.665,0.985)--(2.665,1.136);
\draw[gp path] (2.669,0.985)--(2.669,1.136);
\draw[gp path] (2.672,0.985)--(2.672,1.287);
\draw[gp path] (2.676,0.985)--(2.676,1.136);
\draw[gp path] (2.680,0.985)--(2.680,1.136);
\draw[gp path] (2.687,0.985)--(2.687,1.136);
\draw[gp path] (2.694,0.985)--(2.694,1.287);
\draw[gp path] (2.698,0.985)--(2.698,1.136);
\draw[gp path] (2.702,0.985)--(2.702,1.438);
\draw[gp path] (2.705,0.985)--(2.705,1.136);
\draw[gp path] (2.709,0.985)--(2.709,1.136);
\draw[gp path] (2.717,0.985)--(2.717,1.136);
\draw[gp path] (2.724,0.985)--(2.724,1.438);
\draw[gp path] (2.731,0.985)--(2.731,1.438);
\draw[gp path] (2.735,0.985)--(2.735,1.136);
\draw[gp path] (2.739,0.985)--(2.739,1.287);
\draw[gp path] (2.746,0.985)--(2.746,1.589);
\draw[gp path] (2.750,0.985)--(2.750,1.136);
\draw[gp path] (2.753,0.985)--(2.753,1.287);
\draw[gp path] (2.757,0.985)--(2.757,1.136);
\draw[gp path] (2.761,0.985)--(2.761,1.136);
\draw[gp path] (2.765,0.985)--(2.765,1.287);
\draw[gp path] (2.768,0.985)--(2.768,1.136);
\draw[gp path] (2.772,0.985)--(2.772,1.287);
\draw[gp path] (2.776,0.985)--(2.776,1.136);
\draw[gp path] (2.779,0.985)--(2.779,1.287);
\draw[gp path] (2.790,0.985)--(2.790,1.589);
\draw[gp path] (2.794,0.985)--(2.794,1.136);
\draw[gp path] (2.798,0.985)--(2.798,1.136);
\draw[gp path] (2.801,0.985)--(2.801,1.136);
\draw[gp path] (2.805,0.985)--(2.805,1.438);
\draw[gp path] (2.809,0.985)--(2.809,1.136);
\draw[gp path] (2.812,0.985)--(2.812,1.136);
\draw[gp path] (2.816,0.985)--(2.816,1.136);
\draw[gp path] (2.820,0.985)--(2.820,1.136);
\draw[gp path] (2.824,0.985)--(2.824,1.136);
\draw[gp path] (2.827,0.985)--(2.827,1.287);
\draw[gp path] (2.831,0.985)--(2.831,1.136);
\draw[gp path] (2.838,0.985)--(2.838,1.136);
\draw[gp path] (2.842,0.985)--(2.842,1.136);
\draw[gp path] (2.846,0.985)--(2.846,1.438);
\draw[gp path] (2.849,0.985)--(2.849,1.136);
\draw[gp path] (2.853,0.985)--(2.853,1.136);
\draw[gp path] (2.857,0.985)--(2.857,1.136);
\draw[gp path] (2.860,0.985)--(2.860,1.136);
\draw[gp path] (2.864,0.985)--(2.864,1.136);
\draw[gp path] (2.872,0.985)--(2.872,1.136);
\draw[gp path] (2.879,0.985)--(2.879,1.136);
\draw[gp path] (2.883,0.985)--(2.883,1.136);
\draw[gp path] (2.886,0.985)--(2.886,1.136);
\draw[gp path] (2.894,0.985)--(2.894,1.136);
\draw[gp path] (2.897,0.985)--(2.897,1.136);
\draw[gp path] (2.905,0.985)--(2.905,1.287);
\draw[gp path] (2.908,0.985)--(2.908,1.136);
\draw[gp path] (2.912,0.985)--(2.912,1.287);
\draw[gp path] (2.916,0.985)--(2.916,1.136);
\draw[gp path] (2.949,0.985)--(2.949,1.287);
\draw[gp path] (2.956,0.985)--(2.956,1.136);
\draw[gp path] (2.960,0.985)--(2.960,1.136);
\draw[gp path] (2.968,0.985)--(2.968,1.136);
\draw[gp path] (2.971,0.985)--(2.971,1.287);
\draw[gp path] (2.979,0.985)--(2.979,1.136);
\draw[gp path] (2.982,0.985)--(2.982,1.136);
\draw[gp path] (2.986,0.985)--(2.986,1.136);
\draw[gp path] (3.001,0.985)--(3.001,1.287);
\draw[gp path] (3.004,0.985)--(3.004,1.287);
\draw[gp path] (3.012,0.985)--(3.012,1.136);
\draw[gp path] (3.023,0.985)--(3.023,1.136);
\draw[gp path] (3.027,0.985)--(3.027,1.136);
\draw[gp path] (3.030,0.985)--(3.030,1.287);
\draw[gp path] (3.034,0.985)--(3.034,1.136);
\draw[gp path] (3.038,0.985)--(3.038,1.438);
\draw[gp path] (3.045,0.985)--(3.045,1.136);
\draw[gp path] (3.060,0.985)--(3.060,1.287);
\draw[gp path] (3.064,0.985)--(3.064,1.287);
\draw[gp path] (3.078,0.985)--(3.078,1.136);
\draw[gp path] (3.086,0.985)--(3.086,1.136);
\draw[gp path] (3.115,0.985)--(3.115,1.136);
\draw[gp path] (3.123,0.985)--(3.123,1.136);
\draw[gp path] (3.126,0.985)--(3.126,1.287);
\draw[gp path] (3.130,0.985)--(3.130,1.136);
\draw[gp path] (3.134,0.985)--(3.134,1.136);
\draw[gp path] (3.137,0.985)--(3.137,1.136);
\draw[gp path] (3.145,0.985)--(3.145,1.136);
\draw[gp path] (3.152,0.985)--(3.152,1.136);
\draw[gp path] (3.160,0.985)--(3.160,1.136);
\draw[gp path] (3.171,0.985)--(3.171,1.136);
\draw[gp path] (3.185,0.985)--(3.185,1.438);
\draw[gp path] (3.189,0.985)--(3.189,1.136);
\draw[gp path] (3.196,0.985)--(3.196,1.136);
\draw[gp path] (3.200,0.985)--(3.200,1.287);
\draw[gp path] (3.204,0.985)--(3.204,1.287);
\draw[gp path] (3.215,0.985)--(3.215,1.136);
\draw[gp path] (3.222,0.985)--(3.222,1.136);
\draw[gp path] (3.230,0.985)--(3.230,1.136);
\draw[gp path] (3.233,0.985)--(3.233,1.287);
\draw[gp path] (3.237,0.985)--(3.237,1.136);
\draw[gp path] (3.252,0.985)--(3.252,1.438);
\draw[gp path] (3.256,0.985)--(3.256,1.136);
\draw[gp path] (3.270,0.985)--(3.270,1.136);
\draw[gp path] (3.274,0.985)--(3.274,1.136);
\draw[gp path] (3.285,0.985)--(3.285,1.438);
\draw[gp path] (3.300,0.985)--(3.300,1.136);
\draw[gp path] (3.304,0.985)--(3.304,1.287);
\draw[gp path] (3.311,0.985)--(3.311,1.287);
\draw[gp path] (3.315,0.985)--(3.315,1.287);
\draw[gp path] (3.318,0.985)--(3.318,1.136);
\draw[gp path] (3.322,0.985)--(3.322,1.438);
\draw[gp path] (3.326,0.985)--(3.326,1.287);
\draw[gp path] (3.333,0.985)--(3.333,1.287);
\draw[gp path] (3.337,0.985)--(3.337,1.136);
\draw[gp path] (3.340,0.985)--(3.340,1.287);
\draw[gp path] (3.344,0.985)--(3.344,1.136);
\draw[gp path] (3.348,0.985)--(3.348,1.136);
\draw[gp path] (3.352,0.985)--(3.352,1.136);
\draw[gp path] (3.355,0.985)--(3.355,1.287);
\draw[gp path] (3.359,0.985)--(3.359,1.287);
\draw[gp path] (3.363,0.985)--(3.363,1.136);
\draw[gp path] (3.366,0.985)--(3.366,1.589);
\draw[gp path] (3.370,0.985)--(3.370,1.136);
\draw[gp path] (3.374,0.985)--(3.374,1.136);
\draw[gp path] (3.377,0.985)--(3.377,1.136);
\draw[gp path] (3.381,0.985)--(3.381,1.136);
\draw[gp path] (3.388,0.985)--(3.388,1.287);
\draw[gp path] (3.396,0.985)--(3.396,1.287);
\draw[gp path] (3.403,0.985)--(3.403,1.136);
\draw[gp path] (3.407,0.985)--(3.407,1.136);
\draw[gp path] (3.414,0.985)--(3.414,1.136);
\draw[gp path] (3.418,0.985)--(3.418,1.136);
\draw[gp path] (3.422,0.985)--(3.422,1.287);
\draw[gp path] (3.425,0.985)--(3.425,1.438);
\draw[gp path] (3.433,0.985)--(3.433,1.136);
\draw[gp path] (3.436,0.985)--(3.436,1.287);
\draw[gp path] (3.440,0.985)--(3.440,1.136);
\draw[gp path] (3.448,0.985)--(3.448,1.136);
\draw[gp path] (3.451,0.985)--(3.451,1.287);
\draw[gp path] (3.455,0.985)--(3.455,1.136);
\draw[gp path] (3.459,0.985)--(3.459,1.136);
\draw[gp path] (3.462,0.985)--(3.462,1.287);
\draw[gp path] (3.470,0.985)--(3.470,1.136);
\draw[gp path] (3.473,0.985)--(3.473,1.136);
\draw[gp path] (3.481,0.985)--(3.481,1.136);
\draw[gp path] (3.484,0.985)--(3.484,1.287);
\draw[gp path] (3.492,0.985)--(3.492,1.287);
\draw[gp path] (3.499,0.985)--(3.499,1.438);
\draw[gp path] (3.507,0.985)--(3.507,1.287);
\draw[gp path] (3.510,0.985)--(3.510,1.438);
\draw[gp path] (3.514,0.985)--(3.514,1.136);
\draw[gp path] (3.518,0.985)--(3.518,1.589);
\draw[gp path] (3.521,0.985)--(3.521,1.287);
\draw[gp path] (3.532,0.985)--(3.532,1.136);
\draw[gp path] (3.536,0.985)--(3.536,1.287);
\draw[gp path] (3.540,0.985)--(3.540,1.136);
\draw[gp path] (3.544,0.985)--(3.544,1.136);
\draw[gp path] (3.547,0.985)--(3.547,1.287);
\draw[gp path] (3.555,0.985)--(3.555,1.287);
\draw[gp path] (3.562,0.985)--(3.562,1.136);
\draw[gp path] (3.566,0.985)--(3.566,1.136);
\draw[gp path] (3.573,0.985)--(3.573,1.136);
\draw[gp path] (3.577,0.985)--(3.577,1.136);
\draw[gp path] (3.580,0.985)--(3.580,1.287);
\draw[gp path] (3.584,0.985)--(3.584,1.136);
\draw[gp path] (3.588,0.985)--(3.588,1.136);
\draw[gp path] (3.599,0.985)--(3.599,1.136);
\draw[gp path] (3.614,0.985)--(3.614,1.287);
\draw[gp path] (3.617,0.985)--(3.617,1.136);
\draw[gp path] (3.621,0.985)--(3.621,1.287);
\draw[gp path] (3.625,0.985)--(3.625,1.136);
\draw[gp path] (3.628,0.985)--(3.628,1.136);
\draw[gp path] (3.632,0.985)--(3.632,1.287);
\draw[gp path] (3.640,0.985)--(3.640,1.287);
\draw[gp path] (3.643,0.985)--(3.643,1.287);
\draw[gp path] (3.651,0.985)--(3.651,1.438);
\draw[gp path] (3.658,0.985)--(3.658,1.136);
\draw[gp path] (3.662,0.985)--(3.662,1.287);
\draw[gp path] (3.665,0.985)--(3.665,1.136);
\draw[gp path] (3.669,0.985)--(3.669,1.589);
\draw[gp path] (3.676,0.985)--(3.676,1.136);
\draw[gp path] (3.680,0.985)--(3.680,1.287);
\draw[gp path] (3.684,0.985)--(3.684,1.136);
\draw[gp path] (3.699,0.985)--(3.699,1.136);
\draw[gp path] (3.710,0.985)--(3.710,1.136);
\draw[gp path] (3.713,0.985)--(3.713,1.136);
\draw[gp path] (3.728,0.985)--(3.728,1.136);
\draw[gp path] (3.732,0.985)--(3.732,1.287);
\draw[gp path] (3.739,0.985)--(3.739,1.589);
\draw[gp path] (3.743,0.985)--(3.743,1.287);
\draw[gp path] (3.747,0.985)--(3.747,1.136);
\draw[gp path] (3.750,0.985)--(3.750,1.438);
\draw[gp path] (3.754,0.985)--(3.754,1.136);
\draw[gp path] (3.758,0.985)--(3.758,1.136);
\draw[gp path] (3.761,0.985)--(3.761,1.287);
\draw[gp path] (3.765,0.985)--(3.765,1.438);
\draw[gp path] (3.772,0.985)--(3.772,1.136);
\draw[gp path] (3.780,0.985)--(3.780,1.287);
\draw[gp path] (3.783,0.985)--(3.783,1.287);
\draw[gp path] (3.787,0.985)--(3.787,1.136);
\draw[gp path] (3.791,0.985)--(3.791,1.136);
\draw[gp path] (3.795,0.985)--(3.795,1.136);
\draw[gp path] (3.798,0.985)--(3.798,1.438);
\draw[gp path] (3.802,0.985)--(3.802,1.136);
\draw[gp path] (3.817,0.985)--(3.817,1.287);
\draw[gp path] (3.820,0.985)--(3.820,1.287);
\draw[gp path] (3.824,0.985)--(3.824,1.136);
\draw[gp path] (3.828,0.985)--(3.828,1.438);
\draw[gp path] (3.839,0.985)--(3.839,1.287);
\draw[gp path] (3.843,0.985)--(3.843,1.136);
\draw[gp path] (3.850,0.985)--(3.850,1.136);
\draw[gp path] (3.854,0.985)--(3.854,1.287);
\draw[gp path] (3.857,0.985)--(3.857,1.287);
\draw[gp path] (3.861,0.985)--(3.861,1.287);
\draw[gp path] (3.865,0.985)--(3.865,1.438);
\draw[gp path] (3.868,0.985)--(3.868,1.136);
\draw[gp path] (3.872,0.985)--(3.872,1.136);
\draw[gp path] (3.876,0.985)--(3.876,1.136);
\draw[gp path] (3.879,0.985)--(3.879,1.136);
\draw[gp path] (3.883,0.985)--(3.883,1.136);
\draw[gp path] (3.887,0.985)--(3.887,1.287);
\draw[gp path] (3.891,0.985)--(3.891,1.136);
\draw[gp path] (3.894,0.985)--(3.894,1.287);
\draw[gp path] (3.905,0.985)--(3.905,1.136);
\draw[gp path] (3.909,0.985)--(3.909,1.136);
\draw[gp path] (3.913,0.985)--(3.913,1.136);
\draw[gp path] (3.927,0.985)--(3.927,1.136);
\draw[gp path] (3.935,0.985)--(3.935,1.589);
\draw[gp path] (3.939,0.985)--(3.939,1.287);
\draw[gp path] (3.942,0.985)--(3.942,1.136);
\draw[gp path] (3.953,0.985)--(3.953,1.287);
\draw[gp path] (3.957,0.985)--(3.957,1.438);
\draw[gp path] (3.968,0.985)--(3.968,1.136);
\draw[gp path] (3.972,0.985)--(3.972,1.287);
\draw[gp path] (3.975,0.985)--(3.975,1.287);
\draw[gp path] (3.979,0.985)--(3.979,1.287);
\draw[gp path] (3.983,0.985)--(3.983,1.136);
\draw[gp path] (3.994,0.985)--(3.994,1.136);
\draw[gp path] (3.998,0.985)--(3.998,1.438);
\draw[gp path] (4.023,0.985)--(4.023,1.136);
\draw[gp path] (4.031,0.985)--(4.031,1.589);
\draw[gp path] (4.035,0.985)--(4.035,1.136);
\draw[gp path] (4.038,0.985)--(4.038,1.136);
\draw[gp path] (4.042,0.985)--(4.042,1.136);
\draw[gp path] (4.049,0.985)--(4.049,1.287);
\draw[gp path] (4.053,0.985)--(4.053,1.136);
\draw[gp path] (4.057,0.985)--(4.057,1.136);
\draw[gp path] (4.060,0.985)--(4.060,1.136);
\draw[gp path] (4.068,0.985)--(4.068,1.136);
\draw[gp path] (4.079,0.985)--(4.079,1.136);
\draw[gp path] (4.086,0.985)--(4.086,1.136);
\draw[gp path] (4.090,0.985)--(4.090,1.136);
\draw[gp path] (4.101,0.985)--(4.101,1.287);
\draw[gp path] (4.105,0.985)--(4.105,1.136);
\draw[gp path] (4.108,0.985)--(4.108,1.136);
\draw[gp path] (4.116,0.985)--(4.116,1.287);
\draw[gp path] (4.119,0.985)--(4.119,1.287);
\draw[gp path] (4.127,0.985)--(4.127,1.589);
\draw[gp path] (4.131,0.985)--(4.131,1.136);
\draw[gp path] (4.134,0.985)--(4.134,1.438);
\draw[gp path] (4.138,0.985)--(4.138,1.287);
\draw[gp path] (4.145,0.985)--(4.145,1.136);
\draw[gp path] (4.149,0.985)--(4.149,1.136);
\draw[gp path] (4.156,0.985)--(4.156,1.136);
\draw[gp path] (4.164,0.985)--(4.164,1.438);
\draw[gp path] (4.182,0.985)--(4.182,1.136);
\draw[gp path] (4.186,0.985)--(4.186,1.136);
\draw[gp path] (4.190,0.985)--(4.190,1.287);
\draw[gp path] (4.197,0.985)--(4.197,1.136);
\draw[gp path] (4.204,0.985)--(4.204,1.136);
\draw[gp path] (4.208,0.985)--(4.208,1.287);
\draw[gp path] (4.215,0.985)--(4.215,1.287);
\draw[gp path] (4.219,0.985)--(4.219,1.438);
\draw[gp path] (4.227,0.985)--(4.227,1.136);
\draw[gp path] (4.230,0.985)--(4.230,1.136);
\draw[gp path] (4.238,0.985)--(4.238,1.136);
\draw[gp path] (4.249,0.985)--(4.249,1.287);
\draw[gp path] (4.256,0.985)--(4.256,1.136);
\draw[gp path] (4.263,0.985)--(4.263,1.136);
\draw[gp path] (4.271,0.985)--(4.271,1.287);
\draw[gp path] (4.278,0.985)--(4.278,1.136);
\draw[gp path] (4.289,0.985)--(4.289,1.136);
\draw[gp path] (4.297,0.985)--(4.297,1.287);
\draw[gp path] (4.304,0.985)--(4.304,1.136);
\draw[gp path] (4.308,0.985)--(4.308,1.438);
\draw[gp path] (4.315,0.985)--(4.315,1.287);
\draw[gp path] (4.323,0.985)--(4.323,1.438);
\draw[gp path] (4.330,0.985)--(4.330,1.287);
\draw[gp path] (4.334,0.985)--(4.334,1.136);
\draw[gp path] (4.337,0.985)--(4.337,1.287);
\draw[gp path] (4.341,0.985)--(4.341,1.287);
\draw[gp path] (4.345,0.985)--(4.345,1.136);
\draw[gp path] (4.356,0.985)--(4.356,1.136);
\draw[gp path] (4.367,0.985)--(4.367,1.136);
\draw[gp path] (4.371,0.985)--(4.371,1.136);
\draw[gp path] (4.374,0.985)--(4.374,1.287);
\draw[gp path] (4.378,0.985)--(4.378,1.589);
\draw[gp path] (4.385,0.985)--(4.385,1.287);
\draw[gp path] (4.393,0.985)--(4.393,1.741);
\draw[gp path] (4.396,0.985)--(4.396,1.136);
\draw[gp path] (4.400,0.985)--(4.400,1.136);
\draw[gp path] (4.404,0.985)--(4.404,1.438);
\draw[gp path] (4.407,0.985)--(4.407,1.287);
\draw[gp path] (4.411,0.985)--(4.411,1.287);
\draw[gp path] (4.419,0.985)--(4.419,1.136);
\draw[gp path] (4.422,0.985)--(4.422,1.136);
\draw[gp path] (4.426,0.985)--(4.426,1.136);
\draw[gp path] (4.437,0.985)--(4.437,1.136);
\draw[gp path] (4.441,0.985)--(4.441,1.136);
\draw[gp path] (4.444,0.985)--(4.444,1.287);
\draw[gp path] (4.448,0.985)--(4.448,1.136);
\draw[gp path] (4.452,0.985)--(4.452,1.287);
\draw[gp path] (4.455,0.985)--(4.455,1.287);
\draw[gp path] (4.459,0.985)--(4.459,1.136);
\draw[gp path] (4.463,0.985)--(4.463,1.136);
\draw[gp path] (4.467,0.985)--(4.467,1.287);
\draw[gp path] (4.470,0.985)--(4.470,1.136);
\draw[gp path] (4.474,0.985)--(4.474,1.136);
\draw[gp path] (4.478,0.985)--(4.478,1.287);
\draw[gp path] (4.481,0.985)--(4.481,1.438);
\draw[gp path] (4.485,0.985)--(4.485,1.892);
\draw[gp path] (4.489,0.985)--(4.489,1.287);
\draw[gp path] (4.492,0.985)--(4.492,1.136);
\draw[gp path] (4.496,0.985)--(4.496,1.136);
\draw[gp path] (4.500,0.985)--(4.500,1.136);
\draw[gp path] (4.507,0.985)--(4.507,1.136);
\draw[gp path] (4.511,0.985)--(4.511,1.136);
\draw[gp path] (4.522,0.985)--(4.522,1.438);
\draw[gp path] (4.529,0.985)--(4.529,1.136);
\draw[gp path] (4.533,0.985)--(4.533,1.287);
\draw[gp path] (4.540,0.985)--(4.540,1.136);
\draw[gp path] (4.544,0.985)--(4.544,1.136);
\draw[gp path] (4.548,0.985)--(4.548,1.287);
\draw[gp path] (4.555,0.985)--(4.555,1.136);
\draw[gp path] (4.563,0.985)--(4.563,1.589);
\draw[gp path] (4.574,0.985)--(4.574,1.136);
\draw[gp path] (4.577,0.985)--(4.577,1.287);
\draw[gp path] (4.581,0.985)--(4.581,1.287);
\draw[gp path] (4.585,0.985)--(4.585,1.136);
\draw[gp path] (4.588,0.985)--(4.588,1.136);
\draw[gp path] (4.592,0.985)--(4.592,1.438);
\draw[gp path] (4.596,0.985)--(4.596,1.136);
\draw[gp path] (4.607,0.985)--(4.607,1.136);
\draw[gp path] (4.618,0.985)--(4.618,1.287);
\draw[gp path] (4.622,0.985)--(4.622,1.136);
\draw[gp path] (4.633,0.985)--(4.633,1.287);
\draw[gp path] (4.636,0.985)--(4.636,1.438);
\draw[gp path] (4.640,0.985)--(4.640,1.287);
\draw[gp path] (4.644,0.985)--(4.644,1.136);
\draw[gp path] (4.651,0.985)--(4.651,1.438);
\draw[gp path] (4.655,0.985)--(4.655,1.438);
\draw[gp path] (4.658,0.985)--(4.658,1.136);
\draw[gp path] (4.662,0.985)--(4.662,1.589);
\draw[gp path] (4.666,0.985)--(4.666,1.136);
\draw[gp path] (4.670,0.985)--(4.670,1.287);
\draw[gp path] (4.673,0.985)--(4.673,1.136);
\draw[gp path] (4.677,0.985)--(4.677,1.438);
\draw[gp path] (4.681,0.985)--(4.681,1.287);
\draw[gp path] (4.684,0.985)--(4.684,1.136);
\draw[gp path] (4.695,0.985)--(4.695,1.438);
\draw[gp path] (4.699,0.985)--(4.699,1.287);
\draw[gp path] (4.703,0.985)--(4.703,1.136);
\draw[gp path] (4.718,0.985)--(4.718,1.136);
\draw[gp path] (4.725,0.985)--(4.725,1.136);
\draw[gp path] (4.729,0.985)--(4.729,1.136);
\draw[gp path] (4.732,0.985)--(4.732,1.287);
\draw[gp path] (4.743,0.985)--(4.743,1.136);
\draw[gp path] (4.747,0.985)--(4.747,1.136);
\draw[gp path] (4.751,0.985)--(4.751,1.136);
\draw[gp path] (4.754,0.985)--(4.754,1.136);
\draw[gp path] (4.758,0.985)--(4.758,1.136);
\draw[gp path] (4.769,0.985)--(4.769,1.438);
\draw[gp path] (4.773,0.985)--(4.773,1.287);
\draw[gp path] (4.780,0.985)--(4.780,1.287);
\draw[gp path] (4.784,0.985)--(4.784,1.287);
\draw[gp path] (4.788,0.985)--(4.788,1.136);
\draw[gp path] (4.791,0.985)--(4.791,1.287);
\draw[gp path] (4.795,0.985)--(4.795,1.136);
\draw[gp path] (4.799,0.985)--(4.799,1.136);
\draw[gp path] (4.802,0.985)--(4.802,1.438);
\draw[gp path] (4.806,0.985)--(4.806,1.287);
\draw[gp path] (4.810,0.985)--(4.810,1.136);
\draw[gp path] (4.817,0.985)--(4.817,1.136);
\draw[gp path] (4.821,0.985)--(4.821,1.287);
\draw[gp path] (4.825,0.985)--(4.825,1.438);
\draw[gp path] (4.828,0.985)--(4.828,1.136);
\draw[gp path] (4.832,0.985)--(4.832,1.589);
\draw[gp path] (4.836,0.985)--(4.836,1.287);
\draw[gp path] (4.839,0.985)--(4.839,1.287);
\draw[gp path] (4.847,0.985)--(4.847,1.438);
\draw[gp path] (4.854,0.985)--(4.854,1.438);
\draw[gp path] (4.865,0.985)--(4.865,1.136);
\draw[gp path] (4.869,0.985)--(4.869,1.438);
\draw[gp path] (4.880,0.985)--(4.880,1.136);
\draw[gp path] (4.887,0.985)--(4.887,1.287);
\draw[gp path] (4.891,0.985)--(4.891,1.438);
\draw[gp path] (4.895,0.985)--(4.895,1.136);
\draw[gp path] (4.902,0.985)--(4.902,1.136);
\draw[gp path] (4.910,0.985)--(4.910,1.136);
\draw[gp path] (4.913,0.985)--(4.913,1.287);
\draw[gp path] (4.917,0.985)--(4.917,1.136);
\draw[gp path] (4.928,0.985)--(4.928,1.287);
\draw[gp path] (4.935,0.985)--(4.935,1.438);
\draw[gp path] (4.939,0.985)--(4.939,1.287);
\draw[gp path] (4.943,0.985)--(4.943,1.136);
\draw[gp path] (4.950,0.985)--(4.950,1.287);
\draw[gp path] (4.958,0.985)--(4.958,1.136);
\draw[gp path] (4.961,0.985)--(4.961,1.136);
\draw[gp path] (4.976,0.985)--(4.976,1.136);
\draw[gp path] (4.998,0.985)--(4.998,1.287);
\draw[gp path] (5.002,0.985)--(5.002,1.136);
\draw[gp path] (5.006,0.985)--(5.006,1.287);
\draw[gp path] (5.009,0.985)--(5.009,1.136);
\draw[gp path] (5.013,0.985)--(5.013,1.136);
\draw[gp path] (5.017,0.985)--(5.017,1.136);
\draw[gp path] (5.020,0.985)--(5.020,1.287);
\draw[gp path] (5.028,0.985)--(5.028,1.287);
\draw[gp path] (5.035,0.985)--(5.035,1.136);
\draw[gp path] (5.039,0.985)--(5.039,1.438);
\draw[gp path] (5.046,0.985)--(5.046,1.287);
\draw[gp path] (5.050,0.985)--(5.050,1.136);
\draw[gp path] (5.054,0.985)--(5.054,1.136);
\draw[gp path] (5.057,0.985)--(5.057,1.287);
\draw[gp path] (5.061,0.985)--(5.061,1.136);
\draw[gp path] (5.068,0.985)--(5.068,1.136);
\gpcolor{gp lt color border}
\node[gp node right] at (3.604,3.365) {$\rho^+$};
\gpcolor{gp lt color 1}
\gpsetlinetype{gp lt plot 1}
\draw[gp path] (3.788,3.365)--(4.704,3.365);
\draw[gp path] (1.380,2.345)--(1.417,2.345)--(1.455,2.345)--(1.492,2.345)--(1.529,2.345)%
  --(1.566,2.345)--(1.604,2.345)--(1.641,2.345)--(1.678,2.345)--(1.715,2.345)--(1.753,2.345)%
  --(1.790,2.345)--(1.827,2.345)--(1.864,2.345)--(1.902,2.345)--(1.939,2.345)--(1.976,2.345)%
  --(2.013,2.345)--(2.051,2.345)--(2.088,2.345)--(2.125,2.345)--(2.162,2.345)--(2.200,2.345)%
  --(2.237,2.345)--(2.274,2.345)--(2.311,2.345)--(2.349,2.345)--(2.386,2.345)--(2.423,2.345)%
  --(2.460,2.345)--(2.498,2.345)--(2.535,2.345)--(2.572,2.345)--(2.609,2.345)--(2.647,2.345)%
  --(2.684,2.345)--(2.721,2.345)--(2.758,2.345)--(2.796,2.345)--(2.833,2.345)--(2.870,2.345)%
  --(2.907,2.345)--(2.945,2.345)--(2.982,2.345)--(3.019,2.345)--(3.057,2.345)--(3.094,2.345)%
  --(3.131,2.345)--(3.168,2.345)--(3.206,2.345)--(3.243,2.345)--(3.280,2.345)--(3.317,2.345)%
  --(3.355,2.345)--(3.392,2.345)--(3.429,2.345)--(3.466,2.345)--(3.504,2.345)--(3.541,2.345)%
  --(3.578,2.345)--(3.615,2.345)--(3.653,2.345)--(3.690,2.345)--(3.727,2.345)--(3.764,2.345)%
  --(3.802,2.345)--(3.839,2.345)--(3.876,2.345)--(3.913,2.345)--(3.951,2.345)--(3.988,2.345)%
  --(4.025,2.345)--(4.062,2.345)--(4.100,2.345)--(4.137,2.345)--(4.174,2.345)--(4.211,2.345)%
  --(4.249,2.345)--(4.286,2.345)--(4.323,2.345)--(4.360,2.345)--(4.398,2.345)--(4.435,2.345)%
  --(4.472,2.345)--(4.509,2.345)--(4.547,2.345)--(4.584,2.345)--(4.621,2.345)--(4.658,2.345)%
  --(4.696,2.345)--(4.733,2.345)--(4.770,2.345)--(4.808,2.345)--(4.845,2.345)--(4.882,2.345)%
  --(4.919,2.345)--(4.957,2.345)--(4.994,2.345)--(5.031,2.345)--(5.068,2.345);
\gpcolor{gp lt color border}
\node[gp node right] at (3.604,3.057) {$\rho$};
\gpcolor{gp lt color 2}
\gpsetlinetype{gp lt plot 2}
\draw[gp path] (3.788,3.057)--(4.704,3.057);
\draw[gp path] (1.380,3.659)--(1.384,3.590)--(1.387,3.220)--(1.391,3.310)--(1.395,3.218)%
  --(1.398,3.242)--(1.402,3.203)--(1.406,3.168)--(1.410,3.176)--(1.413,3.179)--(1.417,3.153)%
  --(1.421,3.109)--(1.424,3.034)--(1.428,3.051)--(1.432,2.999)--(1.435,2.980)--(1.439,2.931)%
  --(1.443,2.936)--(1.446,2.951)--(1.450,2.940)--(1.454,2.964)--(1.458,2.995)--(1.461,3.063)%
  --(1.465,2.946)--(1.469,2.995)--(1.472,3.004)--(1.476,3.041)--(1.480,2.989)--(1.483,3.029)%
  --(1.487,2.969)--(1.491,2.942)--(1.494,2.883)--(1.498,2.853)--(1.502,2.866)--(1.506,2.865)%
  --(1.509,2.967)--(1.513,2.899)--(1.517,2.930)--(1.520,2.887)--(1.524,2.898)--(1.528,2.809)%
  --(1.531,2.901)--(1.535,2.753)--(1.539,2.812)--(1.542,2.735)--(1.546,2.705)--(1.550,2.697)%
  --(1.554,2.797)--(1.557,2.827)--(1.561,2.822)--(1.565,2.872)--(1.568,2.822)--(1.572,2.851)%
  --(1.576,2.792)--(1.579,2.765)--(1.583,2.709)--(1.587,2.762)--(1.590,2.692)--(1.594,2.649)%
  --(1.598,2.587)--(1.602,2.519)--(1.605,2.508)--(1.609,2.579)--(1.613,2.581)--(1.616,2.555)%
  --(1.620,2.590)--(1.624,2.600)--(1.627,2.643)--(1.631,2.618)--(1.635,2.714)--(1.638,2.593)%
  --(1.642,2.605)--(1.646,2.591)--(1.650,2.630)--(1.653,2.537)--(1.657,2.620)--(1.661,2.546)%
  --(1.664,2.620)--(1.668,2.573)--(1.672,2.658)--(1.675,2.552)--(1.679,2.490)--(1.683,2.478)%
  --(1.686,2.504)--(1.690,2.528)--(1.694,2.634)--(1.698,2.605)--(1.701,2.606)--(1.705,2.582)%
  --(1.709,2.511)--(1.712,2.504)--(1.716,2.491)--(1.720,2.544)--(1.723,2.473)--(1.727,2.494)%
  --(1.731,2.470)--(1.734,2.436)--(1.738,2.433)--(1.742,2.464)--(1.746,2.455)--(1.749,2.436)%
  --(1.753,2.434)--(1.757,2.493)--(1.760,2.502)--(1.764,2.525)--(1.768,2.534)--(1.771,2.537)%
  --(1.775,2.552)--(1.779,2.526)--(1.782,2.581)--(1.786,2.615)--(1.790,2.558)--(1.794,2.572)%
  --(1.797,2.541)--(1.801,2.596)--(1.805,2.537)--(1.808,2.615)--(1.812,2.582)--(1.816,2.572)%
  --(1.819,2.640)--(1.823,2.615)--(1.827,2.594)--(1.830,2.670)--(1.834,2.664)--(1.838,2.673)%
  --(1.842,2.794)--(1.845,2.674)--(1.849,2.721)--(1.853,2.686)--(1.856,2.800)--(1.860,2.649)%
  --(1.864,2.769)--(1.867,2.727)--(1.871,2.776)--(1.875,2.705)--(1.878,2.754)--(1.882,2.760)%
  --(1.886,2.774)--(1.889,2.759)--(1.893,2.703)--(1.897,2.659)--(1.901,2.686)--(1.904,2.573)%
  --(1.908,2.549)--(1.912,2.493)--(1.915,2.544)--(1.919,2.499)--(1.923,2.559)--(1.926,2.591)%
  --(1.930,2.652)--(1.934,2.572)--(1.937,2.566)--(1.941,2.640)--(1.945,2.661)--(1.949,2.635)%
  --(1.952,2.640)--(1.956,2.519)--(1.960,2.614)--(1.963,2.540)--(1.967,2.594)--(1.971,2.558)%
  --(1.974,2.550)--(1.978,2.513)--(1.982,2.593)--(1.985,2.513)--(1.989,2.546)--(1.993,2.600)%
  --(1.997,2.544)--(2.000,2.546)--(2.004,2.516)--(2.008,2.544)--(2.011,2.549)--(2.015,2.559)%
  --(2.019,2.637)--(2.022,2.573)--(2.026,2.593)--(2.030,2.569)--(2.033,2.606)--(2.037,2.581)%
  --(2.041,2.578)--(2.045,2.644)--(2.048,2.602)--(2.052,2.573)--(2.056,2.590)--(2.059,2.544)%
  --(2.063,2.606)--(2.067,2.572)--(2.070,2.623)--(2.074,2.591)--(2.078,2.529)--(2.081,2.437)%
  --(2.085,2.508)--(2.089,2.451)--(2.093,2.493)--(2.096,2.466)--(2.100,2.458)--(2.104,2.401)%
  --(2.107,2.365)--(2.111,2.354)--(2.115,2.348)--(2.118,2.312)--(2.122,2.268)--(2.126,2.218)%
  --(2.129,2.244)--(2.133,2.233)--(2.137,2.250)--(2.141,2.247)--(2.144,2.287)--(2.148,2.247)%
  --(2.152,2.300)--(2.155,2.324)--(2.159,2.381)--(2.163,2.425)--(2.166,2.387)--(2.170,2.387)%
  --(2.174,2.420)--(2.177,2.378)--(2.181,2.351)--(2.185,2.346)--(2.189,2.333)--(2.192,2.331)%
  --(2.196,2.352)--(2.200,2.300)--(2.203,2.355)--(2.207,2.298)--(2.211,2.291)--(2.214,2.319)%
  --(2.218,2.292)--(2.222,2.313)--(2.225,2.315)--(2.229,2.322)--(2.233,2.327)--(2.237,2.283)%
  --(2.240,2.365)--(2.244,2.337)--(2.248,2.390)--(2.251,2.416)--(2.255,2.434)--(2.259,2.493)%
  --(2.262,2.478)--(2.266,2.455)--(2.270,2.398)--(2.273,2.411)--(2.277,2.454)--(2.281,2.413)%
  --(2.285,2.419)--(2.288,2.325)--(2.292,2.380)--(2.296,2.343)--(2.299,2.410)--(2.303,2.371)%
  --(2.307,2.396)--(2.310,2.383)--(2.314,2.426)--(2.318,2.436)--(2.321,2.499)--(2.325,2.455)%
  --(2.329,2.463)--(2.333,2.443)--(2.336,2.454)--(2.340,2.455)--(2.344,2.482)--(2.347,2.433)%
  --(2.351,2.460)--(2.355,2.436)--(2.358,2.425)--(2.362,2.493)--(2.366,2.513)--(2.369,2.514)%
  --(2.373,2.490)--(2.377,2.488)--(2.381,2.490)--(2.384,2.549)--(2.388,2.537)--(2.392,2.534)%
  --(2.395,2.652)--(2.399,2.627)--(2.403,2.664)--(2.406,2.575)--(2.410,2.558)--(2.414,2.516)%
  --(2.417,2.590)--(2.421,2.513)--(2.425,2.501)--(2.429,2.540)--(2.432,2.535)--(2.436,2.501)%
  --(2.440,2.501)--(2.443,2.540)--(2.447,2.540)--(2.451,2.535)--(2.454,2.482)--(2.458,2.448)%
  --(2.462,2.466)--(2.465,2.446)--(2.469,2.529)--(2.473,2.526)--(2.477,2.569)--(2.480,2.558)%
  --(2.484,2.553)--(2.488,2.514)--(2.491,2.567)--(2.495,2.630)--(2.499,2.606)--(2.502,2.590)%
  --(2.506,2.640)--(2.510,2.603)--(2.513,2.611)--(2.517,2.635)--(2.521,2.641)--(2.525,2.670)%
  --(2.528,2.597)--(2.532,2.552)--(2.536,2.544)--(2.539,2.596)--(2.543,2.578)--(2.547,2.507)%
  --(2.550,2.529)--(2.554,2.541)--(2.558,2.455)--(2.561,2.371)--(2.565,2.340)--(2.569,2.351)%
  --(2.573,2.381)--(2.576,2.330)--(2.580,2.417)--(2.584,2.428)--(2.587,2.464)--(2.591,2.422)%
  --(2.595,2.387)--(2.598,2.355)--(2.602,2.337)--(2.606,2.313)--(2.609,2.245)--(2.613,2.260)%
  --(2.617,2.269)--(2.621,2.269)--(2.624,2.232)--(2.628,2.194)--(2.632,2.212)--(2.635,2.201)%
  --(2.639,2.165)--(2.643,2.254)--(2.646,2.176)--(2.650,2.165)--(2.654,2.141)--(2.657,2.191)%
  --(2.661,2.219)--(2.665,2.213)--(2.669,2.179)--(2.672,2.210)--(2.676,2.177)--(2.680,2.257)%
  --(2.683,2.239)--(2.687,2.210)--(2.691,2.179)--(2.694,2.247)--(2.698,2.170)--(2.702,2.167)%
  --(2.705,2.153)--(2.709,2.142)--(2.713,2.105)--(2.717,2.068)--(2.720,2.082)--(2.724,2.105)%
  --(2.728,2.083)--(2.731,2.088)--(2.735,2.014)--(2.739,2.017)--(2.742,2.062)--(2.746,2.106)%
  --(2.750,2.093)--(2.753,2.120)--(2.757,2.091)--(2.761,2.103)--(2.765,2.099)--(2.768,2.068)%
  --(2.772,2.056)--(2.776,2.026)--(2.779,2.064)--(2.783,2.088)--(2.787,2.043)--(2.790,2.068)%
  --(2.794,2.028)--(2.798,2.016)--(2.801,2.062)--(2.805,2.040)--(2.809,2.082)--(2.812,2.087)%
  --(2.816,2.120)--(2.820,2.141)--(2.824,2.161)--(2.827,2.099)--(2.831,2.147)--(2.835,2.124)%
  --(2.838,2.153)--(2.842,2.167)--(2.846,2.067)--(2.849,2.094)--(2.853,2.014)--(2.857,2.058)%
  --(2.860,2.003)--(2.864,2.022)--(2.868,2.011)--(2.872,2.022)--(2.875,2.020)--(2.879,1.964)%
  --(2.883,1.935)--(2.886,1.937)--(2.890,1.985)--(2.894,1.919)--(2.897,1.955)--(2.901,1.919)%
  --(2.905,1.905)--(2.908,1.914)--(2.912,1.914)--(2.916,1.981)--(2.920,1.940)--(2.923,2.019)%
  --(2.927,1.987)--(2.931,1.970)--(2.934,2.005)--(2.938,2.037)--(2.942,2.040)--(2.945,2.088)%
  --(2.949,2.082)--(2.953,2.102)--(2.956,2.088)--(2.960,2.080)--(2.964,2.034)--(2.968,2.005)%
  --(2.971,1.997)--(2.975,2.000)--(2.979,2.046)--(2.982,2.049)--(2.986,2.064)--(2.990,2.102)%
  --(2.993,2.067)--(2.997,2.090)--(3.001,2.077)--(3.004,2.087)--(3.008,2.056)--(3.012,2.111)%
  --(3.016,2.071)--(3.019,2.079)--(3.023,2.076)--(3.027,2.100)--(3.030,2.085)--(3.034,2.064)%
  --(3.038,2.058)--(3.041,2.076)--(3.045,2.103)--(3.049,2.065)--(3.052,2.115)--(3.056,2.068)%
  --(3.060,2.068)--(3.064,2.040)--(3.067,2.011)--(3.071,1.969)--(3.075,1.991)--(3.078,1.910)%
  --(3.082,1.966)--(3.086,1.931)--(3.089,2.006)--(3.093,1.978)--(3.097,2.038)--(3.100,2.032)%
  --(3.104,2.061)--(3.108,2.031)--(3.112,2.011)--(3.115,2.012)--(3.119,2.040)--(3.123,2.028)%
  --(3.126,2.022)--(3.130,2.055)--(3.134,2.083)--(3.137,2.061)--(3.141,1.997)--(3.145,2.016)%
  --(3.148,1.988)--(3.152,2.065)--(3.156,2.050)--(3.160,2.085)--(3.163,2.074)--(3.167,2.067)%
  --(3.171,2.068)--(3.174,2.080)--(3.178,2.082)--(3.182,2.114)--(3.185,2.150)--(3.189,2.155)%
  --(3.193,2.080)--(3.196,2.142)--(3.200,2.053)--(3.204,2.059)--(3.208,2.059)--(3.211,2.047)%
  --(3.215,2.074)--(3.219,2.070)--(3.222,2.106)--(3.226,2.082)--(3.230,2.077)--(3.233,2.012)%
  --(3.237,2.055)--(3.241,2.038)--(3.244,2.056)--(3.248,2.058)--(3.252,2.031)--(3.256,2.100)%
  --(3.259,2.088)--(3.263,2.058)--(3.267,2.062)--(3.270,1.978)--(3.274,2.047)--(3.278,2.028)%
  --(3.281,2.040)--(3.285,2.000)--(3.289,2.044)--(3.292,2.017)--(3.296,2.071)--(3.300,2.046)%
  --(3.304,2.052)--(3.307,2.079)--(3.311,2.108)--(3.315,2.156)--(3.318,2.139)--(3.322,2.129)%
  --(3.326,2.204)--(3.329,2.204)--(3.333,2.226)--(3.337,2.212)--(3.340,2.247)--(3.344,2.174)%
  --(3.348,2.248)--(3.352,2.257)--(3.355,2.300)--(3.359,2.313)--(3.363,2.395)--(3.366,2.371)%
  --(3.370,2.298)--(3.374,2.348)--(3.377,2.351)--(3.381,2.390)--(3.385,2.348)--(3.388,2.417)%
  --(3.392,2.337)--(3.396,2.301)--(3.400,2.389)--(3.403,2.322)--(3.407,2.348)--(3.411,2.300)%
  --(3.414,2.219)--(3.418,2.226)--(3.422,2.183)--(3.425,2.260)--(3.429,2.191)--(3.433,2.284)%
  --(3.436,2.262)--(3.440,2.312)--(3.444,2.257)--(3.448,2.318)--(3.451,2.316)--(3.455,2.349)%
  --(3.459,2.259)--(3.462,2.272)--(3.466,2.318)--(3.470,2.247)--(3.473,2.295)--(3.477,2.241)%
  --(3.481,2.275)--(3.484,2.245)--(3.488,2.301)--(3.492,2.328)--(3.496,2.330)--(3.499,2.345)%
  --(3.503,2.315)--(3.507,2.324)--(3.510,2.297)--(3.514,2.291)--(3.518,2.280)--(3.521,2.204)%
  --(3.525,2.287)--(3.529,2.229)--(3.532,2.292)--(3.536,2.230)--(3.540,2.212)--(3.544,2.176)%
  --(3.547,2.194)--(3.551,2.161)--(3.555,2.226)--(3.558,2.227)--(3.562,2.203)--(3.566,2.236)%
  --(3.569,2.168)--(3.573,2.245)--(3.577,2.209)--(3.580,2.219)--(3.584,2.233)--(3.588,2.221)%
  --(3.592,2.284)--(3.595,2.179)--(3.599,2.216)--(3.603,2.271)--(3.606,2.280)--(3.610,2.209)%
  --(3.614,2.156)--(3.617,2.201)--(3.621,2.159)--(3.625,2.159)--(3.628,2.297)--(3.632,2.241)%
  --(3.636,2.274)--(3.640,2.281)--(3.643,2.295)--(3.647,2.343)--(3.651,2.413)--(3.654,2.377)%
  --(3.658,2.398)--(3.662,2.404)--(3.665,2.392)--(3.669,2.420)--(3.673,2.366)--(3.676,2.360)%
  --(3.680,2.313)--(3.684,2.395)--(3.688,2.416)--(3.691,2.472)--(3.695,2.396)--(3.699,2.478)%
  --(3.702,2.466)--(3.706,2.433)--(3.710,2.392)--(3.713,2.416)--(3.717,2.457)--(3.721,2.535)%
  --(3.724,2.426)--(3.728,2.478)--(3.732,2.422)--(3.735,2.490)--(3.739,2.392)--(3.743,2.430)%
  --(3.747,2.306)--(3.750,2.315)--(3.754,2.304)--(3.758,2.230)--(3.761,2.248)--(3.765,2.236)%
  --(3.769,2.272)--(3.772,2.263)--(3.776,2.298)--(3.780,2.242)--(3.783,2.245)--(3.787,2.263)%
  --(3.791,2.239)--(3.795,2.236)--(3.798,2.272)--(3.802,2.209)--(3.806,2.266)--(3.809,2.256)%
  --(3.813,2.251)--(3.817,2.272)--(3.820,2.301)--(3.824,2.313)--(3.828,2.278)--(3.831,2.271)%
  --(3.835,2.300)--(3.839,2.284)--(3.843,2.248)--(3.846,2.248)--(3.850,2.218)--(3.854,2.210)%
  --(3.857,2.226)--(3.861,2.226)--(3.865,2.170)--(3.868,2.176)--(3.872,2.124)--(3.876,2.183)%
  --(3.879,2.088)--(3.883,2.109)--(3.887,2.074)--(3.891,2.016)--(3.894,2.016)--(3.898,1.979)%
  --(3.902,1.994)--(3.905,1.964)--(3.909,2.028)--(3.913,2.005)--(3.916,2.028)--(3.920,2.019)%
  --(3.924,2.096)--(3.927,2.065)--(3.931,2.034)--(3.935,2.070)--(3.939,2.034)--(3.942,2.074)%
  --(3.946,2.077)--(3.950,2.073)--(3.953,2.020)--(3.957,2.032)--(3.961,1.985)--(3.964,1.988)%
  --(3.968,2.005)--(3.972,1.988)--(3.975,2.052)--(3.979,1.990)--(3.983,2.012)--(3.987,2.049)%
  --(3.990,2.077)--(3.994,2.085)--(3.998,2.111)--(4.001,2.106)--(4.005,2.108)--(4.009,2.138)%
  --(4.012,2.151)--(4.016,2.127)--(4.020,2.158)--(4.023,2.124)--(4.027,2.168)--(4.031,2.123)%
  --(4.035,2.213)--(4.038,2.167)--(4.042,2.191)--(4.046,2.148)--(4.049,2.100)--(4.053,2.126)%
  --(4.057,2.139)--(4.060,2.162)--(4.064,2.144)--(4.068,2.156)--(4.071,2.129)--(4.075,2.162)%
  --(4.079,2.182)--(4.083,2.147)--(4.086,2.168)--(4.090,2.139)--(4.094,2.145)--(4.097,2.144)%
  --(4.101,2.162)--(4.105,2.135)--(4.108,2.158)--(4.112,2.144)--(4.116,2.135)--(4.119,2.108)%
  --(4.123,2.076)--(4.127,2.177)--(4.131,2.150)--(4.134,2.179)--(4.138,2.226)--(4.142,2.230)%
  --(4.145,2.242)--(4.149,2.173)--(4.153,2.185)--(4.156,2.216)--(4.160,2.170)--(4.164,2.244)%
  --(4.167,2.254)--(4.171,2.271)--(4.175,2.226)--(4.179,2.284)--(4.182,2.173)--(4.186,2.281)%
  --(4.190,2.223)--(4.193,2.259)--(4.197,2.207)--(4.201,2.238)--(4.204,2.201)--(4.208,2.200)%
  --(4.212,2.173)--(4.215,2.206)--(4.219,2.179)--(4.223,2.188)--(4.227,2.224)--(4.230,2.251)%
  --(4.234,2.230)--(4.238,2.259)--(4.241,2.164)--(4.245,2.090)--(4.249,2.085)--(4.252,2.055)%
  --(4.256,2.019)--(4.260,2.062)--(4.263,1.987)--(4.267,1.994)--(4.271,1.941)--(4.275,1.954)%
  --(4.278,1.946)--(4.282,1.990)--(4.286,1.987)--(4.289,2.019)--(4.293,2.041)--(4.297,2.059)%
  --(4.300,2.087)--(4.304,2.100)--(4.308,2.062)--(4.311,2.115)--(4.315,2.079)--(4.319,2.147)%
  --(4.323,2.144)--(4.326,2.155)--(4.330,2.115)--(4.334,2.079)--(4.337,2.077)--(4.341,2.044)%
  --(4.345,2.099)--(4.348,2.040)--(4.352,2.132)--(4.356,2.077)--(4.359,2.173)--(4.363,2.147)%
  --(4.367,2.179)--(4.371,2.195)--(4.374,2.159)--(4.378,2.171)--(4.382,2.165)--(4.385,2.138)%
  --(4.389,2.195)--(4.393,2.117)--(4.396,2.112)--(4.400,2.079)--(4.404,2.142)--(4.407,2.083)%
  --(4.411,2.115)--(4.415,2.055)--(4.419,2.105)--(4.422,2.079)--(4.426,2.112)--(4.430,2.070)%
  --(4.433,2.151)--(4.437,2.120)--(4.441,2.165)--(4.444,2.147)--(4.448,2.167)--(4.452,2.123)%
  --(4.455,2.117)--(4.459,2.079)--(4.463,2.135)--(4.467,2.115)--(4.470,2.093)--(4.474,2.040)%
  --(4.478,2.038)--(4.481,2.061)--(4.485,2.025)--(4.489,2.056)--(4.492,2.012)--(4.496,2.041)%
  --(4.500,2.083)--(4.503,2.035)--(4.507,2.085)--(4.511,2.076)--(4.515,2.067)--(4.518,2.136)%
  --(4.522,2.139)--(4.526,2.209)--(4.529,2.164)--(4.533,2.153)--(4.537,2.144)--(4.540,2.156)%
  --(4.544,2.177)--(4.548,2.227)--(4.551,2.215)--(4.555,2.295)--(4.559,2.286)--(4.563,2.256)%
  --(4.566,2.281)--(4.570,2.352)--(4.574,2.294)--(4.577,2.316)--(4.581,2.307)--(4.585,2.318)%
  --(4.588,2.346)--(4.592,2.434)--(4.596,2.387)--(4.599,2.318)--(4.603,2.307)--(4.607,2.324)%
  --(4.611,2.271)--(4.614,2.294)--(4.618,2.277)--(4.622,2.221)--(4.625,2.260)--(4.629,2.241)%
  --(4.633,2.248)--(4.636,2.239)--(4.640,2.226)--(4.644,2.209)--(4.647,2.203)--(4.651,2.135)%
  --(4.655,2.130)--(4.658,2.209)--(4.662,2.145)--(4.666,2.148)--(4.670,2.192)--(4.673,2.183)%
  --(4.677,2.247)--(4.681,2.260)--(4.684,2.236)--(4.688,2.283)--(4.692,2.259)--(4.695,2.229)%
  --(4.699,2.333)--(4.703,2.346)--(4.706,2.386)--(4.710,2.467)--(4.714,2.433)--(4.718,2.458)%
  --(4.721,2.482)--(4.725,2.537)--(4.729,2.380)--(4.732,2.357)--(4.736,2.387)--(4.740,2.404)%
  --(4.743,2.380)--(4.747,2.416)--(4.751,2.443)--(4.754,2.377)--(4.758,2.342)--(4.762,2.351)%
  --(4.766,2.315)--(4.769,2.318)--(4.773,2.318)--(4.777,2.322)--(4.780,2.343)--(4.784,2.428)%
  --(4.788,2.322)--(4.791,2.426)--(4.795,2.333)--(4.799,2.354)--(4.802,2.247)--(4.806,2.334)%
  --(4.810,2.291)--(4.814,2.390)--(4.817,2.300)--(4.821,2.369)--(4.825,2.278)--(4.828,2.340)%
  --(4.832,2.289)--(4.836,2.238)--(4.839,2.244)--(4.843,2.275)--(4.847,2.278)--(4.850,2.257)%
  --(4.854,2.251)--(4.858,2.280)--(4.862,2.262)--(4.865,2.309)--(4.869,2.330)--(4.873,2.300)%
  --(4.876,2.284)--(4.880,2.369)--(4.884,2.348)--(4.887,2.277)--(4.891,2.287)--(4.895,2.316)%
  --(4.898,2.283)--(4.902,2.325)--(4.906,2.213)--(4.910,2.218)--(4.913,2.275)--(4.917,2.235)%
  --(4.921,2.244)--(4.924,2.298)--(4.928,2.318)--(4.932,2.236)--(4.935,2.235)--(4.939,2.218)%
  --(4.943,2.250)--(4.946,2.238)--(4.950,2.221)--(4.954,2.287)--(4.958,2.272)--(4.961,2.251)%
  --(4.965,2.292)--(4.969,2.309)--(4.972,2.380)--(4.976,2.378)--(4.980,2.408)--(4.983,2.437)%
  --(4.987,2.414)--(4.991,2.401)--(4.994,2.460)--(4.998,2.390)--(5.002,2.426)--(5.006,2.422)%
  --(5.009,2.411)--(5.013,2.411)--(5.017,2.408)--(5.020,2.381)--(5.024,2.408)--(5.028,2.445)%
  --(5.031,2.494)--(5.035,2.402)--(5.039,2.339)--(5.042,2.399)--(5.046,2.342)--(5.050,2.318)%
  --(5.054,2.348)--(5.057,2.319)--(5.061,2.342)--(5.065,2.328)--(5.068,2.366);
\gpcolor{gp lt color border}
\gpsetlinetype{gp lt border}
\draw[gp path] (1.380,4.007)--(1.380,0.985)--(5.072,0.985)--(5.072,4.007)--cycle;
\gpdefrectangularnode{gp plot 1}{\pgfpoint{1.380cm}{0.985cm}}{\pgfpoint{5.072cm}{4.007cm}}
\end{tikzpicture}

%% file: mlgameoflife.bbl
\begin{thebibliography}{10}
\providecommand{\url}[1]{\texttt{#1}}
\providecommand{\urlprefix}{URL }

\bibitem{Bigbee:2007}
Bigbee, A., Cioffi-Revilla, C., Luke, S.: Replication of sugarscape using
  mason. In: Terano, T., Kita, H., Deguchi, H., Kijima, K. (eds.) Agent-Based
  Approaches in Economic and Social Complex Systems IV, Agent-Based Social
  Systems, vol.~3, pp. 183--190. Springer (2007)

\bibitem{Caillou:2012}
Caillou, P., {Gil-Quijano}, J.: Simanalyzer : Automated description of groups
  dynamics in agent-based simulations. In: Proc. of 11th Int. Conf. on
  Autonomous Agents and Multiagent Systems (AAMAS 2012) (2012)

\bibitem{Chen:2010}
Chen, C.C., Clack, C., Nagl, S.: Identifying multi-level emergent behaviors in
  agent-directed simulations using complex event type specifications.
  Simulation  86(1),  41--51 (2010)

\bibitem{Cossentino:2010}
Cossentino, M., Gaud, N., Hilaire, V., Galland, S., Koukam, A.: Aspecs: an
  agent-oriented software process for engineering complex systems. Autonomous
  Agents and Multi-Agent Systems  20(2),  260--304 (2010)

\bibitem{David:2009}
David, D., Courdier, R.: See emergence as a metaknowledge. a way to reify
  emergent phenomena in multiagent simulations? In: Proceedings of ICAART'09.
  pp. 564--569. Porto, Portugal (2009)

\bibitem{Drogoul:2003a}
Drogoul, A., Vanbergue, D., Meurisse, T.: Multi-agent based simulation: Where
  are the agents? In: Sichman, J., Bousquet, F., Davidsson, P. (eds.)
  Multi-Agent-Based Simulation II, Lecture Notes in Computer Science, vol.
  2581, pp. 1--15. Springer (2003)

\bibitem{Elkies:1998}
Elkies, N.: The still-life density problem and its generalizations. Voronoi's
  Impact on Modern Science, Book I. Proceedings of the Institute of Mathematics
  of the National Academy of Sciences of Ukraine  21,  228--253 (1998)

\bibitem{Fates:2010}
Fat\`{e}s, N., Chevrier, V.: How important are updating schemes in multi-agent
  systems? an illustration on a multi-turmite model. In: Proc. of 9th Int.
  Conf. on Autonomous Agents and Multiagent Systems (AAMAS 2010). vol.~1, pp.
  533--540 (2010)

\bibitem{Fates:2010a}
Fat{\`e}s, N.: Does "life" resist asynchrony? In: Adamatzky, A. (ed.) Game of
  Life Cellular Automata, pp. 257--274. Springer (2010)

\bibitem{Ferber:1996}
Ferber, J., M\"uller, J.P.: Influences and reaction: a model of situated
  multiagent systems. In: 2nd International Conference on Multi-agent systems
  (ICMAS'96). pp. 72--79 (1996)

\bibitem{Gil-Quijano:2012}
Gil-Quijano, J., Louail, T., Hutzler, G.: From biological to urban cells:
  Lessons from three multilevel agent-based models. In: Desai, N., Liu, A.,
  Winikoff, M. (eds.) Principles and Practice of Multi-Agent Systems, Lecture
  Notes in Computer Science, vol. 7057, pp. 620--635. Springer (2012)

\bibitem{Grimm:2012}
Grimm, V., Railsback, S.: Pattern-oriented modelling: a 'multi-scope' for
  predictive systems ecology. Philosophical Transactions of the Royal Society
  B: Biological Sciences  367,  298--310 (2012)

\bibitem{Gutknecht:2000}
Gutknecht, O., Ferber, J.: The madkit agent platform architecture. In: Agents
  Workshop on Infrastructure for Multi-Agent Systems. pp. 48--55 (2000)

\bibitem{Kubera:2008}
Kubera, Y., Mathieu, P., Picault, S.: {Interaction-oriented agent simulations:
  From theory to implementation}. In: Proc. of the18th European Conf. on
  Artificial Intelligence (ECAI'08). pp. 383--387 (2008)

\bibitem{Langton:1990}
Langton, C.: Computation at the edge of chaos: Phase transitions and emergent
  computation. Physica D: Nonlinear Phenomena  42(1-3),  12--37 (1990)

\bibitem{Li:2005}
Li, J., Ge, W., Zhang, J., Kwauk, M.: Multi-scale compromise and multi-level
  correlation in complex systems. Chemical Engineering Research and Design
  83(6),  574 -- 582 (2005)

\bibitem{Maier:1998}
Maier, M.: Architecting principles for system of systems. Systems Engineering
  1(4),  267--284 (1998)

\bibitem{Maus:2008}
Maus, C., John, M., R{\"o}hl, M., Uhrmacher, A.: Hierarchical modeling for
  computational biology. In: Bernardo, M., Degano, P., Zavattaro, G. (eds.)
  Formal Methods for Computational Systems Biology, Lecture Notes in Computer
  Science, vol. 5016, pp. 81--124. Springer (2008)

\bibitem{Michel:2007}
Michel, F.: The {IRM4S} model: the influence/reaction principle for multiagent
  based simulation. In: Proc. of 6th Int. Conf. on Autonomous Agents and
  Multiagent Systems (AAMAS 2007). pp. 1--3 (2007)

\bibitem{Michel:2003}
Michel, F., Goua\"ich, A., Ferber, J.: Weak interaction and strong interaction
  in agent based simulations. In: Multi-Agent-Based Simulation III, Lecture
  Notes in Computer Science, vol. 2927, pp. 43--56. Springer (2003)

\bibitem{Moncion:2010}
Moncion, T., Amar, P., Hutzler, G.: Automatic characterization of emergent
  phenomena in complex systems. Journal of Biological Physics and Chemistry
  10,  16--23 (2010)

\bibitem{Morvan:2012a}
Morvan, G., Dupont, D., Soyez, J.B., Merzouki, R.: Engineering hierarchical
  complex systems: an agent-based approach -- the case of flexible
  manufacturing systems. In: Service Orientation in Holonic and Multi Agent
  Manufacturing Control, Studies in Computational Intelligence, vol. 402.
  Springer (2012)

\bibitem{Morvan:2011}
Morvan, G., Veremme, A., Dupont, D.: {IRM4MLS}: the influence reaction model
  for multi-level simulation. In: Bosse, T., Geller, A., Jonker, C. (eds.)
  Multi-Agent-Based Simulation XI, Lecture Notes in Artificial Intelligence,
  vol. 6532, pp. 16--27. Springer (2011)

\bibitem{Muller:2009}
M\"uller, J.P.: Towards a formal semantics of event-based multi-agent
  simulations. In: David, N., Sichman, J. (eds.) Multi-Agent-Based Simulation
  IX, Lecture Notes in Computer Science, vol. 5269, pp. 110--126. Springer
  Berlin / Heidelberg (2009)

\bibitem{Navarro:2011}
Navarro, L., Flacher, F., Corruble, V.: Dynamic level of detail for large scale
  agent-based urban simulations. In: Proc. of 10th Int. Conf. on Autonomous
  Agents and Multiagent Systems (AAMAS 2011). pp. 701--708 (2011)

\bibitem{Parunak:2011}
Parunak, H.: Pheromones, probabilities and multiple futures. In:
  Multi-Agent-Based Simulation XI, Lecture Notes in Artificial Intelligence,
  vol. 6532, pp. 44--60. Springer (2011)

\bibitem{Parunak:2012}
Parunak, H.: Between agents and mean fields. In: Multi-Agent-Based Simulation
  XII, Lecture Notes in Artificial Intelligence, vol. 7124. Springer (2012)

\bibitem{Parunak:2007}
Parunak, H., Brueckner, S.: Concurrent modeling of alternative worlds with
  polyagents. In: Multi-Agent-Based Simulation VII, pp. 128--141. Lecture Notes
  in Computer Science, Springer (2007)

\bibitem{Picault:2011}
Picault, S., Mathieu, P.: An interaction-oriented model for multi-scale
  simulation. In: Twenty-Second International Joint Conference on Artificial
  Intelligence (2011)

\bibitem{Prevost:2009}
Pr\'evost, G., Bertelle, C.: Detection and reification of emerging dynamical
  ecosystems from interaction networks. In: Complex Systems and
  Self-organization Modelling, Understanding Complex Systems, vol.~39, pp.
  139--161. Springer (2009)

\bibitem{Soyez:2011}
Soyez, J.B., Morvan, G., Merzouki, R., Dupont, D., Kubiak, P.: Multi-agent
  multi-level modeling -- a methodology to simulate complex systems. In:
  Proceedings of the 23rd European Modeling \& Simulation Symposium (2011)

\bibitem{Taillandier:2012}
Taillandier, P., Vo, D.A., Amouroux, E., Drogoul, A.: {GAMA}: A simulation
  platform that integrates geographical information data, agent-based modeling
  and multi-scale control. In: Desai, N., Liu, A., Winikoff, M. (eds.)
  Principles and Practice of Multi-Agent Systems, Lecture Notes in Computer
  Science, vol. 7057, pp. 242--258. Springer (2012)

\bibitem{Uhrmacher:2007}
Uhrmacher, A.M., Ewald, R., John, M., Maus, C., Jeschke, M., Biermann, S.:
  Combining micro and macro-modeling in devs for computational biology. In:
  Proceedings of the 39th conference on Winter simulation. pp. 871--880 (2007)

\bibitem{Vo:2012}
Vo, D.A., Drogoul, A., Zucker, J.D., Ho, T.V.: A modelling language to
  represent and specify emerging structures in agent-based model. In: Desai,
  N., Liu, A., Winikoff, M. (eds.) Principles and Practice of Multi-Agent
  Systems, Lecture Notes in Computer Science, vol. 7057, pp. 212--227. Springer
  (2012)

\bibitem{Zeigler:2000}
Zeigler, B., Kim, T., Praehofer, H.: Theory of Modeling and Simulation.
  Academic Press, 2nd edn. (2000)

\end{thebibliography}
